\documentclass[useAMS,usenatbib]{mn2e}
\usepackage{graphicx}
\usepackage{lmodern}
\usepackage{multirow}
\usepackage{slantsc}

\title[Evolution of Prolate Clouds at H\hspace{1pt}{\textsl{\textsc{ii}}} Boundaries]{Evolution of Prolate Molecular Clouds at \HII\ Boundaries: II. Formation of BRCs of asymmetrical morphology}
\author[T. M. Kinnear, J. Miao, G. J. White, K. Sugitani and S. Goodwin]{T. M. Kinnear$^{1}$\thanks{E-mail:
tk218@kent.ac.uk}, J. Miao$^{1}$, G. J. White$^{2,3}$, K. Sugitani$^{4}$ and S. Goodwin$^{5}$ \\
$^{1}$ Centre for Astrophysics and Planetary Science, School of Physical Sciences, University of Kent, Canterbury, CT2 7NH, England \\
$^{2}$ Department of Physics and Astronomy, The Open University, Milton Keynes MK7 6AA , England \\
$^{3}$ Space Science and Technology Department,CCLRC Rutherford Appleton Laboratory, Oxfordshire OX11 0QX, England \\
$^{4}$ Graduate School of Natural Sciences, Nagoya City University, Mizuho-ku, Nagoya 467-8501, Japan\\
$^{5}$ Department of Physics and Astronomy, The University of Sheffield, Western Bank, Sheffield S10 2TN
}

\newcommand{\msun}{M$_{\odot}$}
\newcommand{\codei}{{\sc Code i}}
\newcommand{\codeii}{{\sc Code ii}}
\newcommand{\deuv}{\ensuremath{d_{\mathrm{EUV}}}}

\newcommand{\htwoden}{cm${}^{-3}$}
\renewcommand{\deg}{\ensuremath{{}^{\circ}}}
\newcommand{\mdeg}{{}^{\circ}}
\newcommand{\HI}{H\hspace{1pt}{\sc i}}
\newcommand{\HII}{H\hspace{1pt}{\sc ii}}

\begin{document}

\date{Accepted 20?? Month ??. Received 20?? Month ??; in original form 20?? Month ??}

\pagerange{\pageref{firstpage}--\pageref{lastpage}} \pubyear{20??}

\maketitle

\label{firstpage}

\begin{abstract}
A systematic investigation on the evolution of a prolate cloud at an \HII\ boundary is conducted using Smoothed Particle Hydrodynamics (SPH) in order to understand the mechanism for a variety of irregular morphological structures found at the boundaries of various \HII\ regions. The prolate molecular clouds in this investigation are set with their semi-major axes at inclinations between 0 and 90\deg\ to a plane parallel ionizing radiation flux. A set of 4 parameters, the number density $n$, the ratio of major to minor axis $\gamma$, the inclination angle $\varphi$ and the incident flux $F_{\mathrm{EUV}}$, are used to define the initial state of the simulated clouds. The dependence of the evolution of a prolate cloud under Radiation Driven Implosion (RDI) on each of the four parameters is investigated. It is found that: i) in addition to the well studied standard type A, B or C Bright Rimmed Clouds (BRCs), many other types such as asymmetrical BRCs, filamentary structures and irregular horse-head structures could also be developed at \HII\ boundaries with only simple initial conditions; ii) the final morphological structures are very sensitive to the 4 initial parameters, especially to the initial density and the inclination; iii) The previously defined ionizing radiation penetration depth can still be used as a good indicator of the final morphology.

Based on the simulation results, the efficiency of the RDI triggered star formation from clouds of different initial conditions is also estimated. Finally a unified mechanism for the various morphological structures found in many different \HII\ boundaries is suggested.

\end{abstract}


\section{Introduction}
As the intensive EUV radiation from newly formed star(s) ionizes the star-facing surface layer of a molecular cloud, the ionization heating induces a shock which propagates into the cloud, compressing material to form highly condensed core(s). These cores are the potential sites for EUV radiation triggered new star formation. Simultaneously the H\hspace{1pt}$\alpha$ emission produced by the recombination of ions with electrons creates a bright rim around the star facing side of the cloud. The structures formed are termed Bright Rimmed Clouds (BRCs), the best candidate for studying the feedback of massive stars to surrounding and parental molecular cloud(s). This process is the Radiation Driven Implosion (RDI) mechanism for EUV radiation triggered star formation in BRCs.

Multi-wavelength observations have revealed various physical and morphological properties of BRCs at \HII\ regions. BRCs observed so far tend to be classified as type A, B and C in increasing order of the curvature of their bright rim \citep{SugitaniEtAl1991-1, SugitaniOgura1994-1}, though some of BRCs are found in M shaped morphology \citep{Osterbrock-1957, UrquhartEtAl2006-1, Karr-2005}.

Starting with a uniform and spherical molecular cloud illuminated by a plane-parallel ionizing radiation from one side, theoretical modelling based on the RDI mechanism successfully revealed the formation process of standard type A, B or C BRCs. These simulated BRCs are all symmetric to their structure axis, and the latter aligns with the radiation flux direction, i.e., the line connecting the tip of a BRC and the centre of the exciting star \citep{Bertoldi1989-1, LeflochLazareff1994-1, KesselBurkert2003-1, MiaoEtAl2006-1, MiaoEtAl2009-1, Gritschneder2009-1, MiaoEtAl2010-1, BisbasEtAl2011-1, HaworthHarries2012-1, Tremblin-2012}.

\begin{figure}
\center
\includegraphics[width=0.47\textwidth]{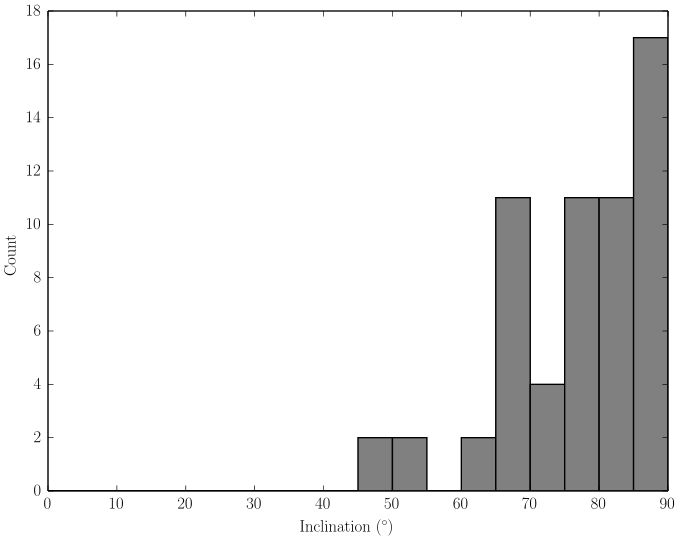}
\caption{The distribution of the inclination angle $\varphi$, which is defined as the angle between the major axis and the $x$ axis as shown in Figure \ref{orientation}, of the BRCs in about 10 emission nebul\ae\ \citep{Osterbrock-1957}.}
\label{angle-distribution}
\end{figure}

However, further investigation on the observed structures at \HII\ boundaries finds that: i) some BRCs do not show a comet-like morphology, but fragment-core linear structures perpendicular to the EUV radiation flux direction (e.g. \citet{ChauhanEtAl2011-1}); ii) most type A, B and C BRCs are not symmetric about their structural axes, which also do not necessarily align with the radiation flux direction of their exciting stars \citep{Thompson-2004, MorganEtAl2004-1}. As shown in Figure \ref{angle-distribution}, there is a distribution in the angle between the axis of a BRC and the direction from the tip of the BRC to the centre of the exciting star, based on the observational data on groups of BRCs in about 10 emission nebul\ae\ \citep{Osterbrock-1957}; iii) There are more objects with structures which can not be categorized by type A, B or C than those which can be; iv) In those asymmetrical BRC structures, RDI triggered multi-star or sequential formation are often found not only at the head of BRCs, but also in the more compressed side layer \citep{FukudaEtAl2013-1, Makela-2013, Sicilia-Aguilar-2014, Panwar-2014}. It is obvious that it is difficult to accommodate the variety of structural and physical features observed with RDI modelling starting with a spherical cloud.

\begin{figure}
\center
\includegraphics[width=0.47\textwidth]{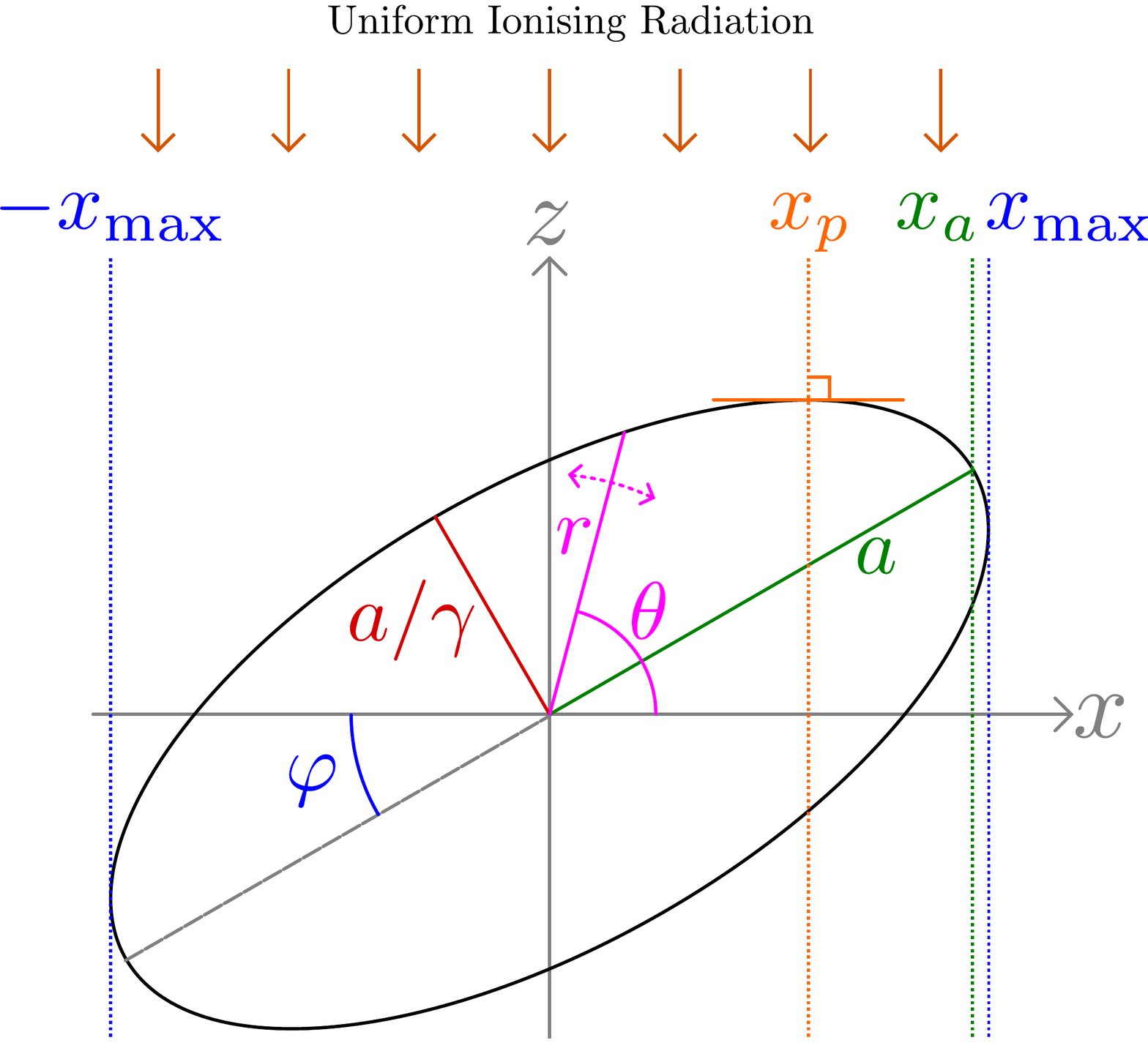}
\caption{The initial geometry of a prolate cloud in simulations. $x_{\mathrm{p}}$, $x_{\mathrm{a}}$ and $x_{\mathrm{max}} $ are the $x$ projections of the apex, the semi-major axis $a$ and the farthest point of the cloud from $z$ axis respectively.}
\label{orientation}
\end{figure}

Observations on the physical properties of molecular clouds have revealed that spherical clouds are very rare cases \citep{Gammie-2003,Doty-2005, RathborneEtAl2009-1} and there are many physical processes which lead to formation of prolate molecular clouds \citep{Hao-2011, Gholipour-2013}. To investigate the mechanism for the development of structures of various morphologies other than standard (i.e., symmetrical) BRCs at \HII\ boundaries, using an initially non-spherical cloud, such as a prolate cloud, in RDI modelling seems the most feasible choice. Definition in terms of two additional, geometric, parameters are introduced: the ratio of the major ($a$) to minor ($b$) axis $\gamma = \frac{a}{b}$ and the inclination angle $\varphi$, the configuration of which is illustrated in Figure \ref{orientation}. In this way we can expect that the variety of the morphological structures derived from RDI simulations must be greatly increased.

A set of RDI simulations were conducted using an initially prolate cloud with $\varphi = 0 \mdeg $, which we call `perpendicular' to the EUV radiation flux. It is found that a perpendicular prolate cloud would evolve into a perpendicular fragment-core linear structure to the ionizing flux, covered by a bright top layer on the star side. Both the initial physical/geometrical conditions of the cloud and the strength of radiation flux affect the distribution of condensed cores over the structure, the total condensed core mass and the triggered star formation time. The sporadic fragment-core structures at \HII\ boundaries can be interpreted by the effect of the interaction between a perpendicular prolate cloud and a parallel-plane EUV radiation field. More details are presented in \citet{Kinnear2014-1}.

In this paper, we expand the investigation to the evolution of a prolate cloud at \HII\ boundaries, with the cloud's major axis inclined to the EUV radiation flux by an angle $ 0\mdeg \le (90\deg - \varphi) \le 90{\mdeg}$. This is performed in order to understand the mechanism for the formation of a broad range of asymmetrical structures and the consequent triggered star formation inside them. In the rest of the paper, we first present a brief description of the codes used, the initial conditions of the clouds for simulations, and a derived quantity which gives an indication of the dynamic evolution of prolate clouds at different \HII\ boundaries. In the results and discussion section we present simulation results and discuss the physical mechanism responsible for the formation of different mophological structures at \HII\ boundaries. Finally, conclusions are drawn.

\section{The codes and initial conditions}
\label{code-initial}
\subsection{The codes}
\label{code}
An updated Smoothed Particles Hydrodynamics (SPH) \codeii\ is used for all of the simulations investigated in this paper, which is an extended version of \codei\ \citep{NelsonLanger1999-1}. \codei\ is an SPH code to investigate the effect of an isotropic FUV interstellar background radiation on the evolution of a molecular cloud, by solving the full set of hydrodynamic equations, including self-gravity of the cloud and a chain of chemical differential equations. Based on \codei, a numerical solver with a ray-tracing algorithm \citep{KesselBurkert2000-1} for a plane-parallel EUV radiation transferring equation was implemented, as well as the relevant heating and cooling processes. We refer to this version as \codeii, which can be used to investigate the interaction of a plane-parallel ionizing radiation from massive star(s) with a molecular cloud of arbitrary initial geometry. More detailed descriptions of the code can be found in \citet{MiaoEtAl2006-1} and \citet{Kinnear2014-1}.

By virtue of the asymmetrical morphology of the simulated clouds, the condensed cores triggered by RDI induced shocks are not usually aligned with the structural axis and multi-star formation is also frequently observed. To identity the positions of these triggered condensed cores and analyse their physical properties, a code called `{\sc CoreFinder}' is used, for which a detailed description is included in \citet{Kinnear2014-1}. For the simulations presented in this paper, all `cores' are defined as a region of at least 0.03 \msun\ ($> 100$ SPH particles) and the density of each SPH particle sampled must be $n \ge 10^6$ \htwoden. As in \citet{Kinnear2014-1}, the cores formed from the simulations are the first generation of RDI triggered objects. Simulations for the subsequent generations of RDI triggered star formation becomes excessively slower after initial extremely high density core(s) form, which leads to a decrease towards infinitesimally small time steps, and the effective halt of the simulation. This can be avoided by the use of the `sink' particles of \citet{BateBurkert1997-1}, but is not currently implemented in \codeii.

All of the column density images are produced using the software {\sc SPLASH}, which is specially designed for processing SPH numerical data \citep{Splash}.

\subsection{Initial conditions}
\label{initil}
\subsubsection{Initial geometry}
The geometrical shape of a prolate cloud in the simulations is described by a pair of initial parameters ( $a$, $\gamma$ ) and its orientation to the plane-parallel ionization radiation flux is $ 0\mdeg \le \varphi \le 90\mdeg$, as shown in Figure \ref{orientation}. The ionization flux onto the surface of the cloud on the star side is treated as plane-parallel to the $z$ axis, for an assumed large ratio of distance to the illuminating star against cloud size.

In Figure \ref{orientation}, $x_{\mathrm{p}}$ is the $x$ coordinate of the apex of the cloud. letting $r$ be the distance of any a point on the surface of a prolate cloud in the $xz$ plane, and $\theta$ the angle between $r$ and the $x$ axis, we have,
\begin{equation}
%r = \frac{a b}{\sqrt{b^2 cos^2 (\theta - \varphi) + a^2 sin^2(\theta - \varphi)}}.
r = \frac{a}{\sqrt{\cos^2 (\theta - \varphi) + \gamma^2 \sin^2(\theta - \varphi)}}.
\end{equation}
The maximum $x$ coordinate and the $x$ projection of the semi-major axis
\begin{eqnarray}
%x_{\mathrm{max}} & = & \sqrt{a^2 \cos^2 \varphi + b^2 \sin^2 \varphi}, \\
x_{\mathrm{max}} & = & a \sqrt{\cos^2 \varphi + \frac{\sin^2 \varphi}{\gamma^2}}, \\
\label{xmax}
x_{\mathrm{a}} & = & a \cos \varphi.
\label{xa}
\end{eqnarray}

As an important indicator to the dynamical evolution of a prolate cloud with an inclination angle $\varphi$, the effective illuminated area by the EUV radiation is,
\begin{equation}
%A = \pi b x_{\mathrm{max}} = \pi b \sqrt{a^2 cos^2 \varphi + b^2 sin^2 \varphi},
A = \frac{\pi a x_{\mathrm{max}}}{\gamma} = \frac{\pi a^2}{\gamma} \sqrt{\cos^2 \varphi + \frac{\sin^2 \varphi}{\gamma^2}},
\label{eff_area}
\end{equation}
which is a decreasing function with $\varphi$, has the maximum $\pi a b$ at $\varphi = 0\mdeg$ and the minimum $\pi b^2$ at $\varphi = 90\mdeg$. For the convenience of the following discussion, we call the half cloud with $\varphi \le \theta \le \pi + \varphi $ as the `front' semi-ellipsoid, and $\pi + \varphi < \theta \le 2 \pi + \varphi $ the `back' semi-ellipsoid.

\subsubsection{Initial mass distribution}
As we have done in the investigation of the evolution of a perpendicular prolate cloud at an \HII\ boundary \citep{Kinnear2014-1}, all of the molecular clouds in our simulations start with a uniform density, which is rendered by a glass-like distribution of SPH particles created using {\sc Gadget-2} \citep{Springel2005-1}. A glass-like distribution is taken as a good approximation to a uniform mass distribution. We choose the initial mass of the cloud to be 30 \msun, the same as used for \citet{Kinnear2014-1}. The number of SPH particles for each molecular cloud is determined by the mass resolution, $3.0 \times 10^{-4}$ \msun\ per SPH particle, for 100,000 particles, higher than required by the convergence test of the \codeii. A zero initial velocity field is set for all of the molecular clouds in the simulations.

Since the objective of our investigation is to explore the evolution of the RDI triggered collapse of an initially inclined prolate cloud, any initially unstable cloud against the self-gravity before interacting with the ionization radiation flux should not be chosen. Stability is assured by applying the Jeans criteria for an isolated prolate cloud, in terms of Jeans number $J$, the ratio of the initial gravitational energy of a prolate cloud to the thermal energy and can be expressed as \citep{Bastien1983-1},
\begin{equation}
J = \frac{\pi \; G \; \rho\; \mu\; b^2}{15\; e\; R_{\mathrm{g}}\; T} \ln{\left(\frac{1 + e }{1 - e }\right)} \le 1.0,
\label{jeansnumber}
\end{equation}
where $\rho$, $T$ and $\mu$ are the mass density, the initial temperature and the mean molecular mass of the prolate cloud respectively, $G$ and $R_{\mathrm{g}}$ are the standard physical constants and the eccentricity,
\begin{equation}
e = \sqrt{1 - \frac{b^2}{a^2} }= \frac{\sqrt{\gamma^2 - 1}}{\gamma}.
\end{equation}
Using the relation between density, mass $M$ and volume of a cloud, we can derive the condition for the major axis $a$ of an initially stable prolate cloud,
\begin{eqnarray}
a \ge a_{\mathrm{crit}} = 0.052 \; \frac{M^*\; \gamma}{T\; \sqrt{\gamma^2 - 1}} \ln \left(\frac{\gamma + \sqrt{\gamma^2-1}}{\gamma - \sqrt{\gamma^2 - 1}}\right),
\label{majoraxis}
\end{eqnarray}
where $M^*$ is the mass of the prolate cloud in units of solar masses, and $a$ and $a_{\mathrm{crit}}$ have units of Parsecs.
For a given molecular cloud of mass $M^*$, and initial temperature $T$ and $\gamma$, a minimum value $a$ can be estimated, the major axis of an initially gravitationally stable cloud should satisfy $a > a_{\mathrm{crit}}$.

\subsection{Boundary condition}

The clouds in our simulations are subject to an isotropic interstellar background FUV radiation of one Habing unit \citep{Habing1968-1} and an ionizing EUV radiation with a flux of $ 10^{9}$ cm$^{-2}$ s$^{-1}$ (typical of the boundary of an \HII\ \ region) directed parallel to the $z$-axis (along the negative $z$ direction) as illustrated in Figure \ref{orientation}.

A constant pressure boundary condition is applied with the value of the external pressure being set equivalent to an external medium of atomic hydrogen of $n\mathrm{(\HI)} = 10 $ \htwoden \ and temperature of 100 K. An outflow condition is imposed at a fixed boundary which is a few times of the initial size of the cloud \citep{NelsonLanger1999-1}.

\subsection{The ionization penetration depth parameter}

%In order to classify the initial dynamic region of an inclined prolate cloud to the ionization radiation field by an angle $\varphi$, we defined the dimensionless ionization penetration parameter as the ratio of the physical ionization penetration depth to minor axis when $\varphi = 0 \mdeg$, $\deuv = \frac{F_{\mathrm{EUV}} \, \gamma }{\alpha_{\mathrm{B}} \, a \, n^2} $ \citep{Kinnear2014-1}.
In order to classify the initial dynamic regime of a perpendicular prolate cloud ($\varphi = 0$), a dimensionless parameter, \deuv, was defined, being the ratio between the physical ionising radiation penetration depth to the semi-minor axis of the cloud. For this scenario, $\deuv = \frac{F_{\mathrm{EUV}} \, \gamma }{\alpha_{\mathrm{B}} \, a \, n^2}$ \citep{Kinnear2014-1}.
When $\varphi \ne 0 \mdeg$, we modify its definition to the ratio of the physical ionization penetration depth to the characteristic depth of the cloud. This characteristic depth is defined as half of the longest path through the cloud in the direction of the radiation, which is always the depth of the cloud at $x=0$, $y=0$.
\begin{eqnarray}
\deuv & = & \frac{\frac{F_{\mathrm{EUV}}}{\alpha_{\mathrm{B}} n^2}}{r(\theta = 90\mdeg)} \nonumber \\
& = & \frac{F_{\mathrm{EUV}} \, \gamma }{\alpha_{\mathrm{B}} \, a \, n^2} \sqrt{\frac{sin^2 \varphi}{\gamma^2} + cos^2 \varphi} \nonumber \\
& = & 1.6 \times 10^3 \, \frac{F^*_{\mathrm{EUV}}\, \gamma}{n^2 \, a^*} \sqrt{\frac{sin^2 \varphi}{\gamma^2} + cos^2 \varphi} \quad (\mathrm{pc})
\label{pene-depth}
\end{eqnarray}
where the major axis $a^{*}$ is in the unit of pc, and $F_{\mathrm{EUV}}^*$ is the EUV ionising radiation flux in units of $10^9$ cm$^{-2}$s$^{-1}$, $\alpha_\mathrm{B}$ is the recombination coefficient of hydrogen ion - election under the `on-the-spot' approximation \citep{DysonWilliams1997-1} and has the value of $2.0 \times 10^{-13}$ cm$^3$ s$^{-1}$ at a temperature of about $10^4$ K \citep{DysonWilliams1997-1}. This is then taken as a constant, as the equilibrium temperature for ionized material is $\approx 10^4$ K and the dependence of $\alpha_\mathrm{B}$ on temperature is not strong in the region around that temperature.

As already discussed in \citet{Kinnear2014-1}, when $\deuv << 1$ for $\varphi = 0$, the prolate cloud is in the RDI triggered shock dominant region, and the collapse of the cloud is by `foci convergence' mode, i.e., two high density cores form 
at the two ends of a filament;  when $\deuv \le 1$, the cloud can still be RDI triggered to collapse but through `linear convergence', i.e., a few high
 density cores form along the whole filament; when $\deuv \ge 1$, the cloud is in photo-ionization dominant region, and would be photo-evaporated. As \deuv\ can possess a similar range of values for all $\varphi$ in Equation \ref{pene-depth}, we would expect it to play a similar role for $\varphi \neq 0$.

\section{Results and discussions}
Three sets of simulations were conducted using prolate clouds of different geometrical shapes ($\gamma$), inclination angles ($\varphi$), initial densities ($n$) and ionising fluxes ($F_{\mathrm{EUV}}$), to investigate the effect of their variations on the evolution of a prolate cloud. Table \ref{seriessummary} contains a summary of the ranges of properties for each set. All of the clouds have an initial temperature of 60 K and are illuminated by a 50,000 K star providing a flux at the cloud, $F_0 = 10^9$ cm$^{-2}$ s$^{-1}$ (or 2.0 $\times 10^9$ cm$^{-2}$ s$^{-1}$, when specified), the same as that for \citet{Kinnear2014-1}. In this section, we first describe the evolution of a prolate cloud of an inclination angle of 45$\mdeg$ in detail, then investigate the influence of changing initial inclination angle $\varphi$ on the evolution of the same cloud. Next we discuss how the initial geometry of a prolate cloud affects the morphological evolution of a cloud and RDI triggered star formation inside it. Finally we propose a formation mechanism for the variety in morphological structures found at \HII\ boundaries.

\begin{table}
\centering
\begin{tabular}{lcccc}
\cline{1-5} 
 & $n$ & $\gamma$ & $\varphi$ & $F_0$ \\
Series name & \htwoden\ & & ${}^{\circ}$ & $\times 10^{9}$ cm$^{-2}$ s$^{-1}$ \\
G1200 & 1200 & 1-10 & 0-90 & 1 \\
density \& flux & 100-1200 & 2 & 60 & 1 \& 2 \\
irregular & 400 & 2-3.5 & 60-85 & 2\\
\cline{1-5}
\end{tabular}
\caption{Summary of the test series conducted. $n$ is initial density, $\gamma$ the axial ratio, $\varphi$ the inclination angle and $F_0$ the incident flux.}
\label{seriessummary}
\end{table}

\begin{table}
\centering
\begin{tabular}{lcccc}
\cline{1-5}
\multicolumn{5}{c}{G1200} \\
No. & $\gamma$ & $a_{\mathrm{crit}}$ (pc) & $a_{1200}$ (pc) & \deuv\ (\%) \\
1 & 1.00 & 0.052 &  0.494  & 0.225 \\
2 & 1.25 & 0.060 &  0.573 &0.218 \\
3 & 1.50 &  0.067 &  0.648  &0.219 \\
4& 1.75 &  0.073&  0.718  &0.221 \\
5 & 2.00 &  0.079& 0.784  & 0.224\\
6 & 2.25 &  0.084&  0.849  & 0.228 \\
7 & 2.50 &   0.039&  0.910  & 0.232\\
8 & 2.75 &  0.093&  0.970  & 0.237\\
9 & 3.00 &  0.097&  1.028   & 0.241 \\
10& 3.25 & 0.101& 1.084    & 0.246\\
11 & 3.50& 0.104& 1.139   &0.250\\
12 & 3.75 &0.107& 1.193  & 0.256\\
13 & 4.00 & 0.110&  1.245 & 0.260\\
14  & 4.50 &0.116& 1.347  & 0.268\\
15 & 5.00 & 0.121& 1.445 & 0.277 \\
16 & 5.50 &0.126& 1.540  &0.285\\
17 & 6.00 &0.130& 1.632  &0.292 \\
18 & 7.00 & 0.138& 1.808 & 0.307\\
19 & 8.00 & 0.145&  1.977 & 0.320\\
\cline{1-5}
\end{tabular}
\caption{Parameters of molecular clouds of G1200 series with an inclination angle $\varphi = 45\mdeg$. From left to right, columns 1-3 are the number identity, axial ratio and the critical semi major axis defined by Equation \ref{majoraxis}. Columns 4-5 are the major axis and \deuv\ defined by Equation \ref{pene-depth}. All of the semi-major axes and critical semi-major axes are in units of pc and the penetration depth is in \%.}
\label{G1200(5)-angle-45}
\end{table}

\subsection{Evolution of cloud G1200(5) with $\varphi = 45\mdeg$}
\label{G1200_5}
Table \ref{G1200(5)-angle-45} lists all of the relevant parameters of clouds with initial density 1200 \htwoden\ and inclination angle $\varphi = 45\mdeg$, for variations of the initial ratio $\gamma$ . The cloud is the fifth one in the G1200 set, notated G1200(5). It has an initial shape defined by $\gamma = 2.0 $, its major axis (0.784 pc) is $\approx 10$ times of its critical major axis (0.079 pc) derived by using Equation \ref{majoraxis}; indicating its initial stability against gravitational collapse.

\begin{figure*}
%\begin{minipage}{0.85\textwidth}
\center
\includegraphics[width=0.85\textwidth]{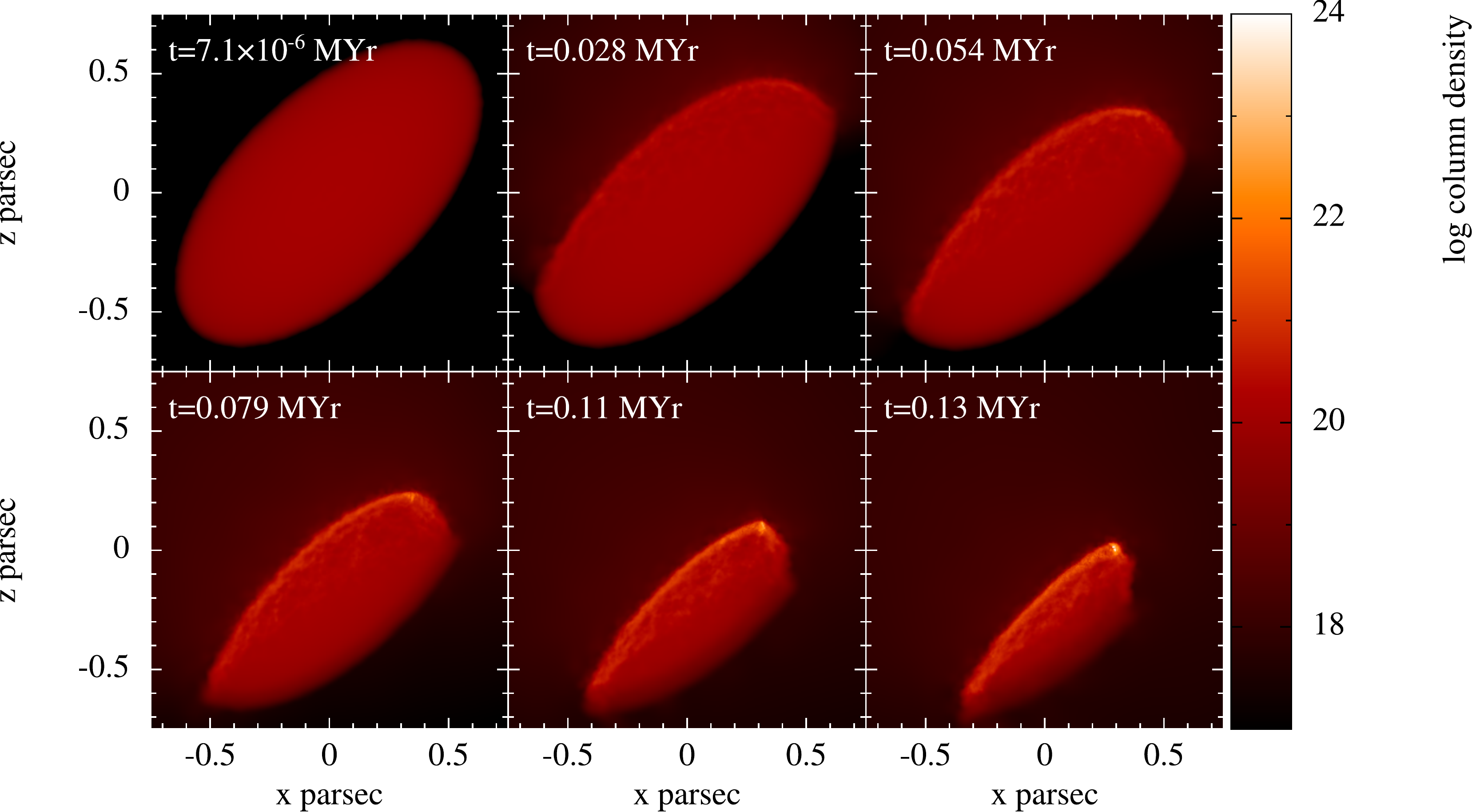}
%\end{minipage}
\caption{Sequential evolution of the column number density for Cloud G1200(5) with the incline angle $\varphi = 45\mdeg$. Time sequence is left to right, then top to bottom then left to right.}
\label{col-den-1200-45}
\end{figure*}

\begin{figure*}
%\begin{minipage}{0.85\textwidth}
\center
\includegraphics[width=0.85\textwidth]{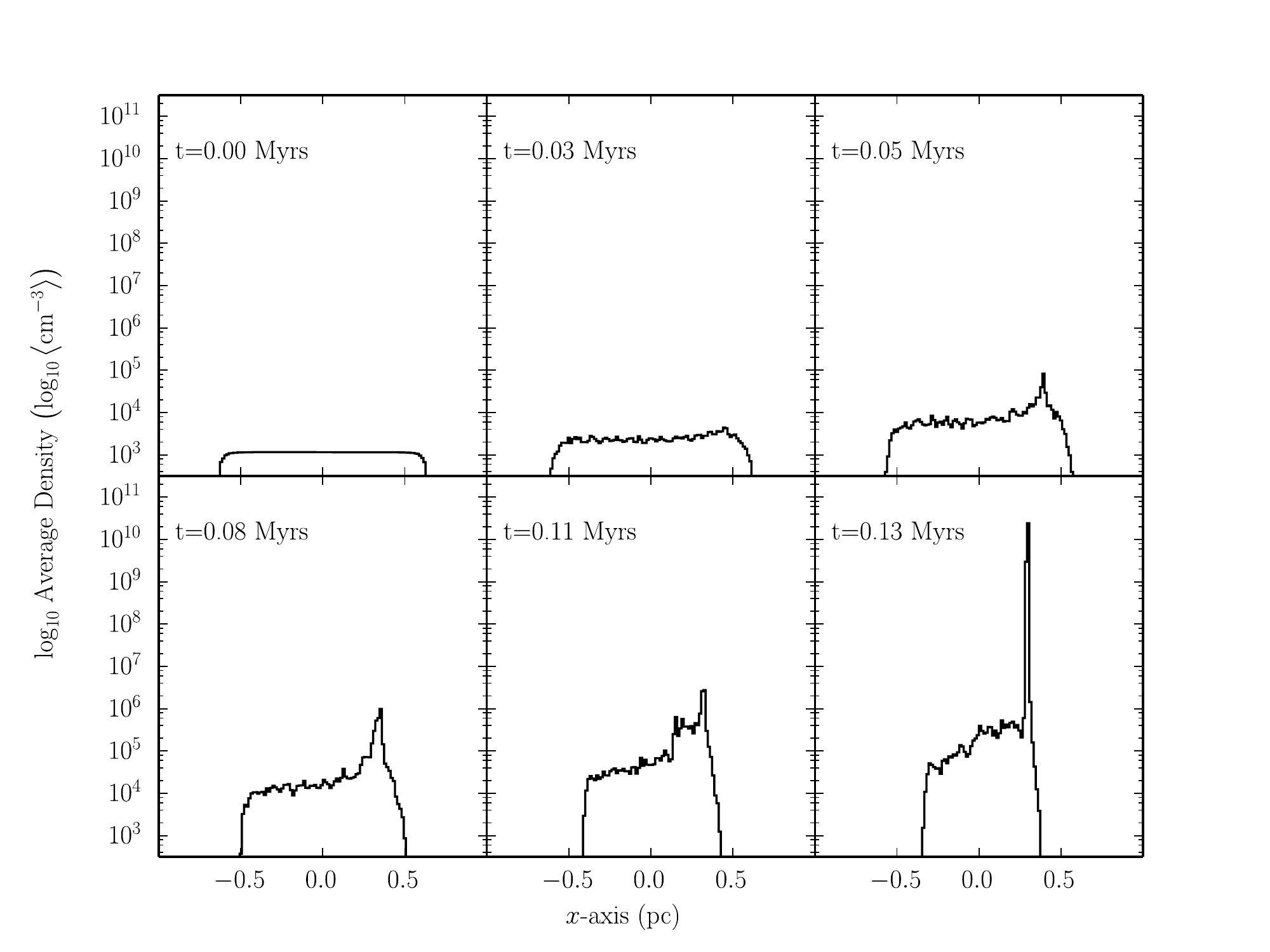}
%\end{minipage}
\caption{Axial mean density distribution along the $x$ axis of the cloud G1200(5) corresponding to the column density evolutionary snapshots in Figure \ref{col-den-1200-45} }
\label{axial-den-1200-45}
\end{figure*}

Figure \ref{col-den-1200-45} shows an evolutionary sequence of column density of the cloud over a period of 0.13 Myr. We also plot the corresponding mean axial density distribution along the $x$ axis in Figure \ref{axial-den-1200-45}, with the established binning method for describing axial profiles of SPH particle properties \citep{Kinnear2014-1, NelsonLanger1999-1, Nelson-1993}. This view provides a qualitative picture of the location of condensed cores formed in the shocked layer of gas in the cloud by the RDI effect.

It is seen from the above two figures that 28 Kyr after the radiation flux was switched on, a shock is established by the ionization heating. This shock starts compressing the cloud surface on the star-facing side. A slightly compressed thin layer of mean density $\approx 3.0 \times 10^3$ \htwoden\ appears at the EUV exposed surface, as shown in the upper-middle panels in Figures \ref{col-den-1200-45} and \ref{axial-den-1200-45}. The density at $x = 0.49$ pc is slightly higher than the axial mean density, which reflects the very early stage of a condensed core formation around the apex of the cloud.

\begin{figure}
\center
\includegraphics[width=0.5\textwidth]{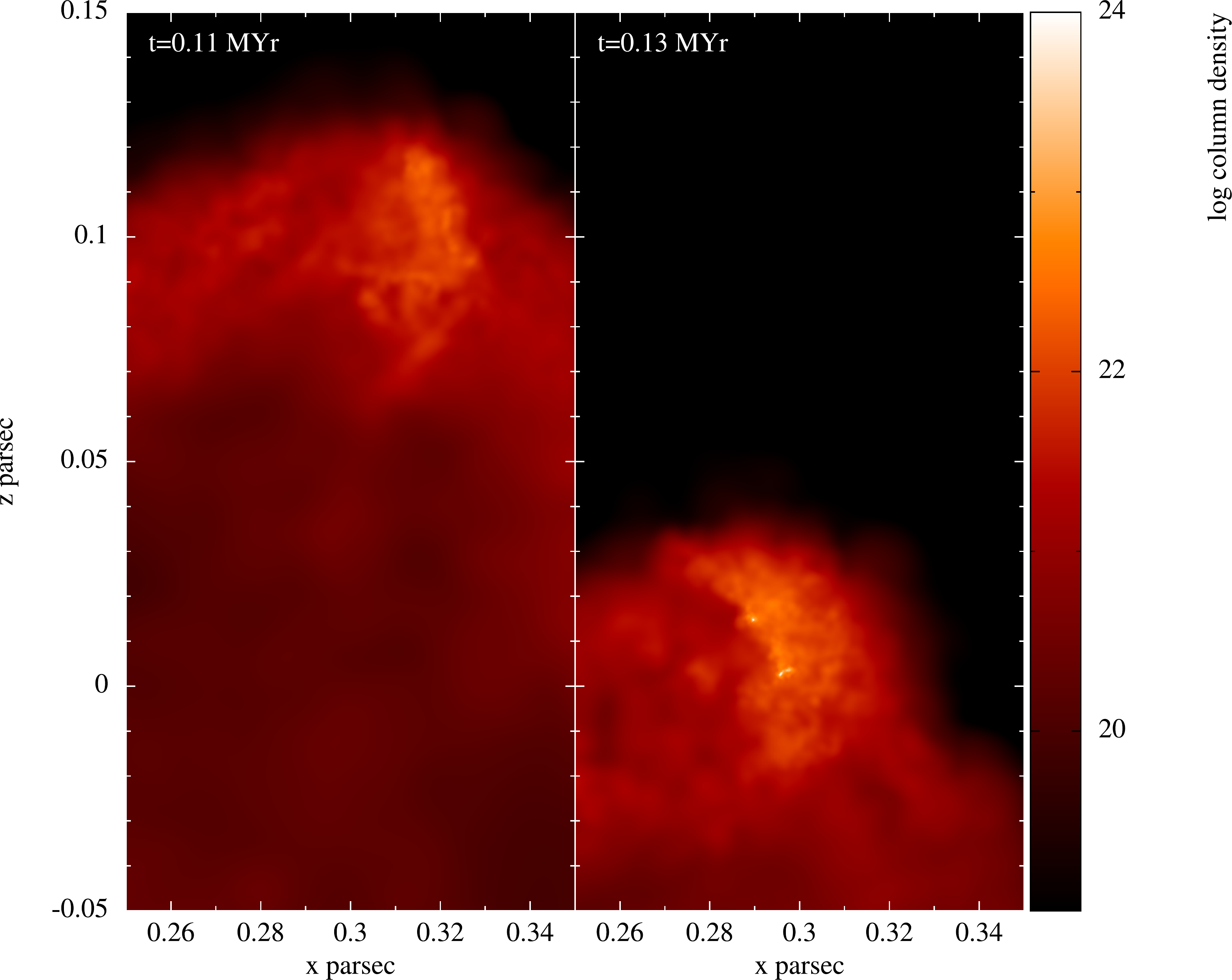}
\caption{Magnification of the column density profiles time $t = 0.11$ (left) and 0.13 Myr (right) at the head of the cloud G1200(5) with $\varphi = 45\mdeg$}
\label{zooming-col-1200-45}
\end{figure}

From $t = 0.028$ to 0.11 Myr, the shock continues propagating into the cloud. The density in the thin shocked layer continues increasing and the structure of the head becomes more distinctive. The non-head mean axial density and the central density of the head increase to $10^4$ and $10^5$ \htwoden\ respectively at 0.054 Myr, and to $2.0 \times 10^4$ and $10^6$ \htwoden\ respectively at 0.08 Myr. At $t = 0.11$ Myr, the density in the shocked layer increases dramatically towards the apex of the cloud (at $x = 0.32$ pc). This density jumps from $\approx 8 \times 10^4$ to $8 \times 10^5$ \htwoden\ at $x = 0. 24$ pc, then to the maximum $4 \times 10^6$ \htwoden\ at the apex. It is interesting to find from the magnified image in the left panel of Figure \ref{zooming-col-1200-45} that a filament structure forms at the head, aligning with the EUV radiation direction. The structure has a length of roughly 0.03 pc and the top end of it coincides with the apex of the cloud.

The gas contained in both the shocked surface layer and in the filament becomes further compressed by $t = 0.13$ Myr. At this stage the maximum mean density in the shocked layer and in the filament are $ 10^5 - 10^6 $ and $ > 10^9$ \htwoden\ respectively.
%Further looking at the axial density profile at $t = 0.13$ Myr shows that there are two high density cores around the apex at $0.32$ pc, with densities being $4 \times 10^9$ and $2 \times 10^{10}$ \htwoden, which might be the consequence of fragmentation of the filament formed at $t = 0.11$ Myr.
The right panel of Figure \ref{zooming-col-1200-45} further reveals the formation of two high density cores in close proximity. These dual cores may be the result of fragmentation of the filament-like structure observed at $t=0.11$ Myr.

Overall, the cloud G1200(5) with an initial inclination angle of 45$\mdeg$ evolves to an asymmetrical type B BRC, with double RDI triggered high density cores embedded at its head, very close to the apex of the BRC. The cores are the potential sites for new star formation. At the same time the apex of the cloud moves down from $z = 0.61$ pc to $z = 0.035$ pc, a net recoil velocity of $\left< v_z \right> = 4.3$ km s$^{-1}$, due to the rocket effect caused by the evaporating gas from the star-facing surface irradiated by the ionizing radiation \citep{OortSpitzer1955-1}. Next we investigate the effect of the inclination of a prolate cloud on its dynamical evolution.

\subsection{Evolution of G1200(5) with $\bf 0\mdeg \le \varphi \le 90 \mdeg $}
To investigate the dependence on the inclination, the cloud G1200(5) is simulated for a range of inclinations $0\mdeg \le \varphi \le 90 \mdeg$. The sequence of inclination angles are chosen as 0, 15, 30, 45, 60, 75 and 90 $\mdeg$. The values of \deuv\ for these seven clouds are listed in Table \ref{G1200(5)-diff-angle}, from which we can see that the ionizing penetration depth decreases with $\varphi$, and is very shallow, with values $<< 1$. As a result, all of these clouds are in the RDI induced shock-dominated regime and triggered collapse is expected.

\begin{table}
\centering
\begin{tabular}{llllllll}
\cline{1-8}
$\varphi (\mdeg)$ & 0 & 15 & 30 & 45 & 60 & 75 & 90 \\
\deuv& \multirow{2}{*}{0.283} & \multirow{2}{*}{0.276} & \multirow{2}{*}{0.255} & \multirow{2}{*}{0.224} & \multirow{2}{*}{0.187} & \multirow{2}{*}{0.155} & \multirow{2}{*}{0.141} \\
(\%) &&&&&&&\\
\cline{1-8}
\end{tabular}
\caption{Variation of \deuv\ over different $\varphi$ for cloud G1200(5).}
\label{G1200(5)-diff-angle}
\end{table}

\subsubsection{Morphological variation with $\varphi$}
\begin{figure*}
%\begin{minipage}{0.85\textwidth}
\includegraphics[width=0.85\textwidth]{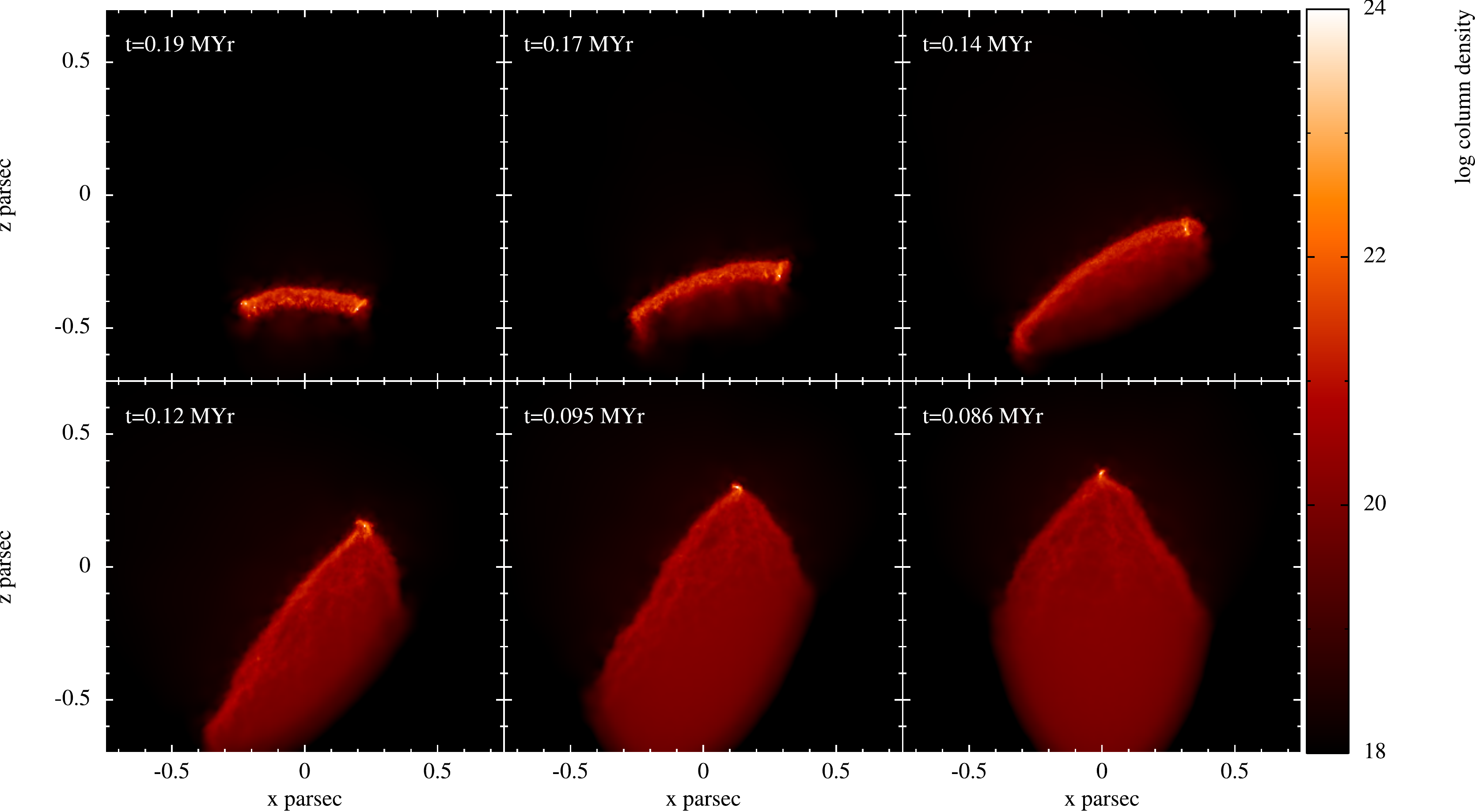}
%\end{minipage}
\caption{The final snapshot of the column density from six simulations of G1200(5) with different initial inclination angles; left to right then up to bottom are per: 0, 15, 30, 60, 75 90 $\mdeg$.}
\label{col-den-1200-all}
\end{figure*}

It is found that the evolutionary sequences of the 7 clouds are qualitatively similar to that with $\varphi = 45\mdeg$, i.e., the formation of a condensed layer on the star facing surface of the cloud and highly condensed core(s) near the apex, although the final morphologies themselves depend on the initial inclination angle of the cloud. As the general evolution is so similar, they are not discussed individually, but with focus on comparison of the differences in their final morphologies and the physical properties of any high density core(s). In Figure \ref{col-den-1200-all}, the final snapshots of column density of cloud G1200(5) from 6 different simulations with $\varphi = 0, 15, 30, 60, 75, $ and $90 \mdeg$, are presented. The equivalent for $\varphi = 45\mdeg$ can be found in Figure \ref{col-den-1200-45}.

When the inclination angle $\varphi $ is small ($ \le 30\mdeg$), $x_a \approx x_{\mathrm{max}}$, the EUV radiation mainly illuminates the surface of the front semi-ellipsoid. The effect of the RDI is to drive a pseudo-plane-parallel shock propagating into the cloud, and the prolate cloud evolves to a filamentary structure with a condensed surface layer on the star side and highly condensed core(s) at the apex (when $\varphi \ne 0\mdeg$) or at either the apex or the two ends (when $\varphi = 0\mdeg$). For the latter case, the evolution has been analysed in detail by \citet{Kinnear2014-1}.
% Two inclined filamentary structure are found in the simulations with initial angles $\varphi = 15$ and $30\mdeg$.%%%%%%%%%%%%%%%%TMK - Why here?

When $ 45 \mdeg \le \varphi < 90\mdeg$, the EUV radiation illuminates not only the surface of the front semi-ellipsoid, but also the top part of back semi-elliptical surface, so that the shock induced by RDI creates a condensed and curved surface layer with the apex as a convergence point of two compressed curving surface layers, those surfaces on opposing sides of the apex. The region around the apex is also the site for the formation of high density core(s). These clouds evolve to asymmetrical BRCs of different morphologies. The morphological sequence is similar to that of the cloud with $\varphi = 45\mdeg$, as shown in Figure \ref{col-den-1200-45}. Comparing the three final morphologies of the simulations at 45, 60 and 75\deg\ in Figures \ref{col-den-1200-45} and \ref{col-den-1200-all}, we can see that as the inclination angle increases, the final structure of the asymmetrical BRC becomes increasingly extended both in the head and tail. This is because, as the effective EUV illumination area decreases as described in Equation \ref{eff_area}, the total volume of the cloud affected by the RDI induced shock becomes smaller. Consequently the original prolate shape of a cloud is less distorted.

When $\varphi = 90\mdeg $, the cloud G1200(5) develops into a symmetrical type B/C BRC but having a wide tail structure, with the widest part defined by the initial semi-minor axis. The morphology developed is different from that of standard type B/C BRC structure, which has a much narrower width in its tail, if one assumes that the progenitor cloud was initially spherical \citep{LeflochLazareff1994-1, KesselBurkert2003-1, MiaoEtAl2006-1, MiaoEtAl2009-1}. Observations have also reported several type B/C BRCs with wide tail morphological structures, e.g. SFO74 \citep{Thompson-2004,Kusune2014-1}.

In summary, a variety of non-standard morphological structures can be obtained by changing the inclination of cloud G1200(5). At low inclination, linear  fragment-core structures form, at middle inclinations, asymmetric type B and C structures, and approaching 90\deg inclinations these return to symmetrical type B and C structures.

\subsubsection{The location of high density cores}
\begin{figure}
\center
\includegraphics[width=0.5\textwidth]{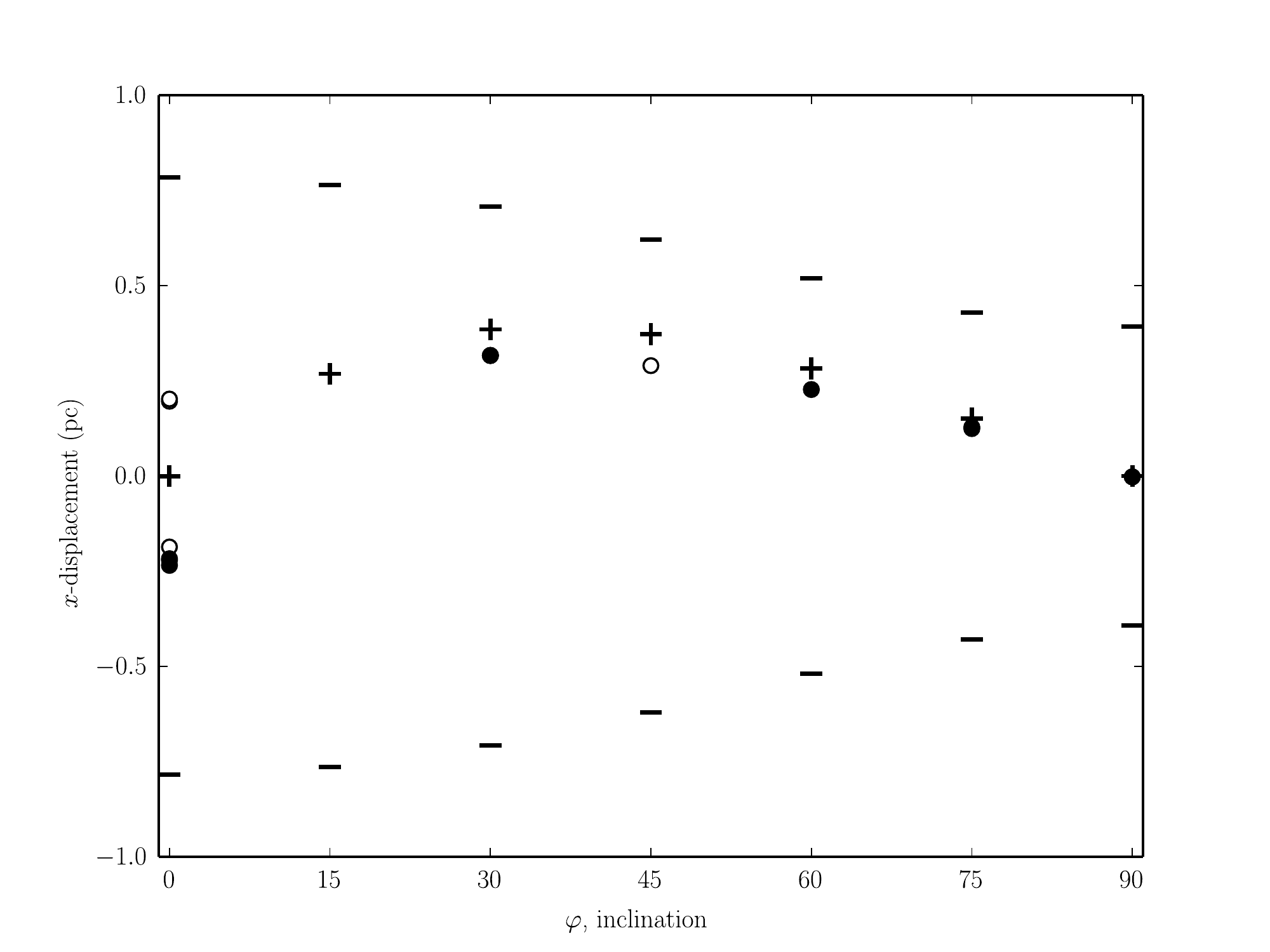}
\caption{The $x$ displacement of the cores formed in G1200(5) with different initial inclination angles. The short `-' indicates $x_{\mathrm{max}}$, `+' $x_{\mathrm{p}}$, `$\circ$' for the cores with $10^6 \le n_{\mathrm{peak}} \le 10^{12}$ \htwoden, and `$\bullet$' for the extremely high density cores with $n_{\mathrm{peak}} > 10^{12}$ \htwoden.}
\label{dis-x-1200-r2-all}
\end{figure}

Figure \ref{dis-x-1200-r2-all} illustrates the distribution of the $x$-displacements of condensed cores formed in the 7 simulations of different initial inclination angle $\varphi$. A point to note is that, where cores form in very close proximity, their positions cannot be distinguished on the scale used, regardless of the size choice of the  marker. It can be seen that when $\varphi > 0\mdeg $, core(s) always form at the side of positive $x$ axis and the change of the $x$ displacement with $\varphi$ follows the trend of $x_{\mathrm{p}}$ with $\varphi$. This illustrates an extremely high preference for high density core(s) to form around the apex $x_p$ in each simulation.

\begin{figure}
\center
\includegraphics[width=0.43\textwidth]{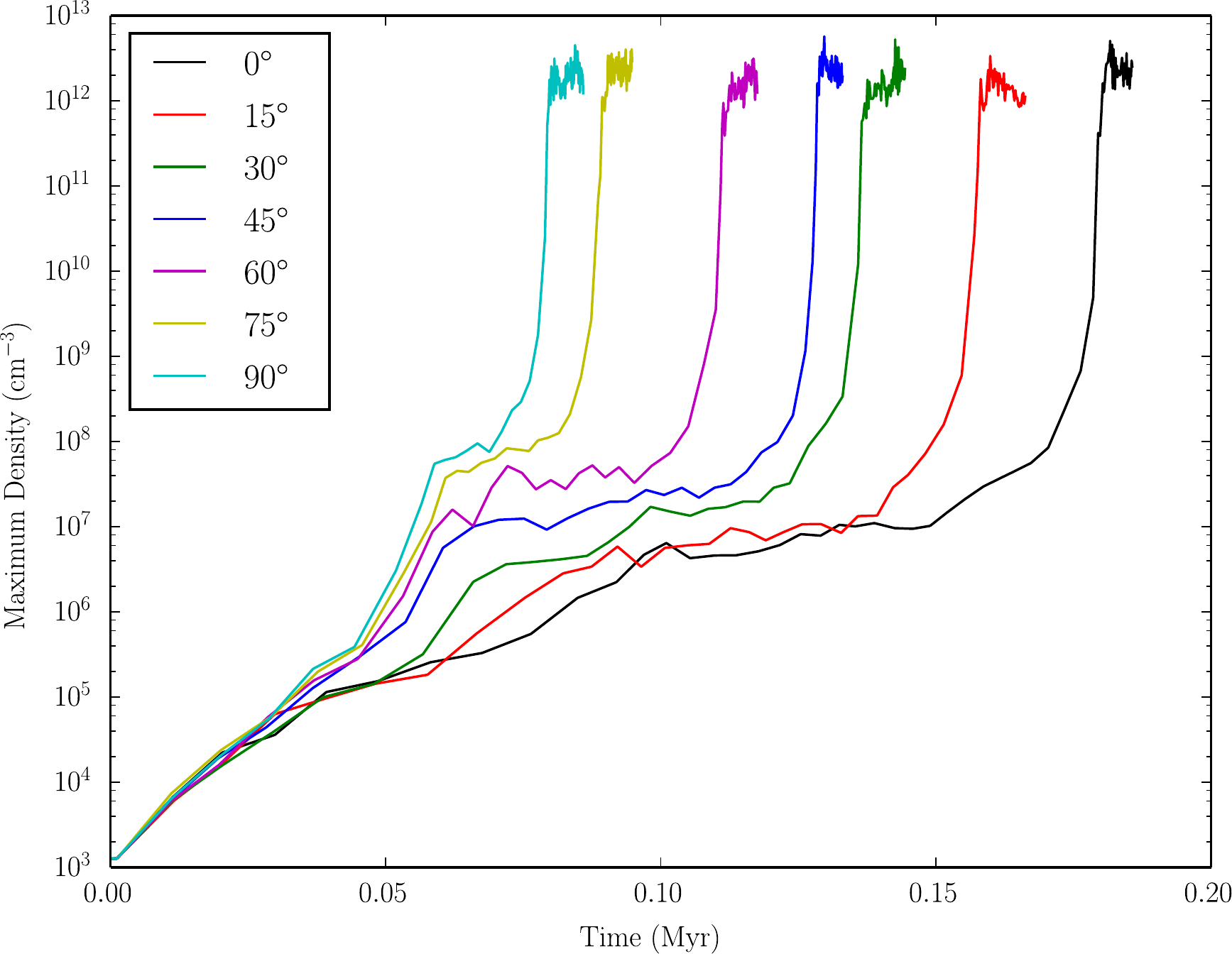}
\caption{The evolution of maximum density of the core(s) in the cloud G1200(5) with different initial inclination angles.}
\label{max_den_with_time_plot}
\end{figure}

\begin{figure}
\center
\includegraphics[width=0.45\textwidth]{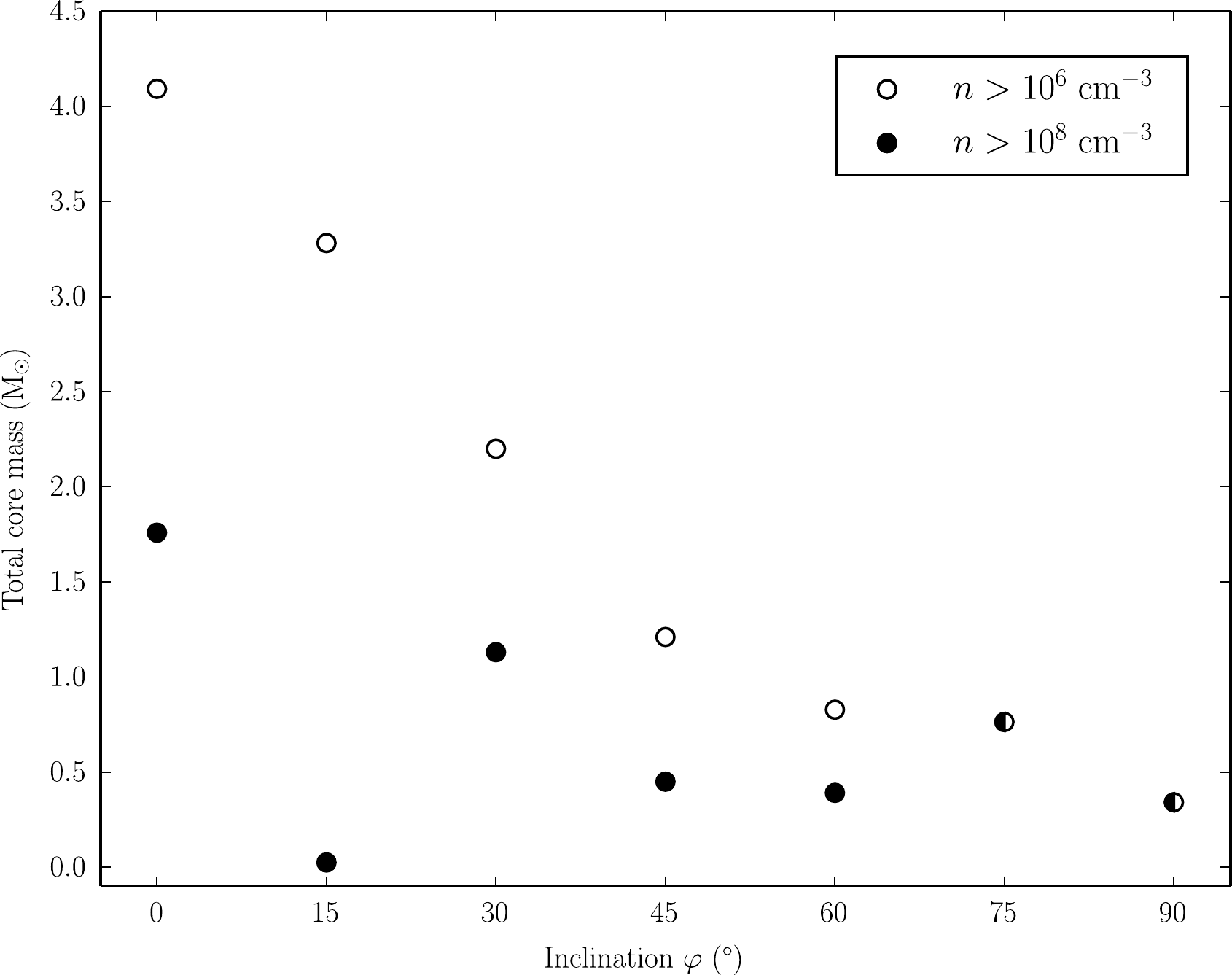}
\caption{The variation of the total core mass of the cloud G1200(5) over different initial inclination angles. The symbols and their represented properties of cores are shown in the legend box.}
\label{total_core_mass_E06-08}
\end{figure}

\subsubsection{RDI triggered core formation efficiency}

In order to examine the effect of varying inclination angle on the efficiency of RDI triggered high density core formation, we investigate the angle dependence of the core formation time and the accumulated core mass. Figure \ref{max_den_with_time_plot} shows the evolution of the maximum density in the 7 simulations with different inclination angles. In general, the evolution of $n_{\mathrm{max}}$ in each simulation passes through three different phases: quasi-linear, quasi-stable and steeply rising, although the time of each phase varies with inclination angle $\varphi$. These phases may correspond to the different stages of RDI triggered high density core formation. Beginning with initial compression by an RDI induced shock, mass accumulates to form the condensed region, which finally collapses to form high density core(s). The time for a simulation to reach the highest density (high density core formation) decreases with inclination angle, from 0.19 down to 0.086 Myr for simulations with $\varphi = 0$ and $90\mdeg$ respectively. This is because the curvature of the EUV radiation illuminated surface around the apex of the prolate cloud increases with $\varphi$, this is beneficial for mass accumulating towards a geometric focus underneath the curved surface. This decreases the path length to the focus, and so also the time for accumulating mass around it and triggering gravitational instability of the condensed region; albeit with less material having been `swept up' by the front. Due to this shortening path length, the total mass of the high density core(s) is therefore expected to decrease with $\varphi$.

The relation between the total high density core mass and inclination angle $\varphi$ is shown in Figure \ref{total_core_mass_E06-08}, which reveals that the total high density ($ n \ge 10^6 $ \htwoden) core mass is about 4.1 \msun, in simulation of $\varphi = 0$, then decreases with $\varphi$, becoming 0.4 \msun\ at $\varphi = 90\mdeg$. This is consistent with the above expectation. However the total mass of the extremely high density core(s) ($n > 10^8$ \htwoden) doesn't show an obvious similar relation with $\varphi$  to the mass of high density core(s).
%This may be due to the different ways in which the extremely dense cores form: by direct gravitational collapsing of a high density region, or collapsing of a fragment from a high density region. The former leads to a single higher mass core and the latter a lower mass core or cores. From the profile of these extremely dense core(s), it seems that the mass of RDI triggered stars is not very strongly dependent on the inclination angle.
This may be for several reasons. Firstly, as the simulation is curtailed before the end of all of the dynamic processes, extremely high density cores may be on the cusp of formation, but have not yet had the opportunity. Secondly, a large degree of random variation is expected with such gravitational instability-based collapse. Accumulating mass in compact high density regions can be fairly consistent, but the collapse of these regions or parts of these regions to extremely high densities through gravitational instability is chaotic and varies strongly with the exact configuration of the dynamics.

From the above result, it is understood that as a cloud is rotated from $\varphi = 0$ to $90\deg$, the morphology of the cloud varies from fragment-core filamentary structures to varied asymmetrical BRCs, then to standard symmetrical BRCs as $\varphi$ approaches $90\deg$. At high $\varphi$, RDI triggered star formation occurs earlier with lower masses at high densities. However, for the sampling performed, such pattern is not obvious for the extremely high density core formation, which appears to be chaotic. Next we look at the effect of the initial shape of an inclined prolate cloud on its final morphological structure and RDI triggered star formation efficiency.

\subsection{Evolution of G1200 series with ${\bf 1.0 \le \gamma \le 8.0}$}
The initial shape of the clouds investigated here is defined by the ratio of the major to minor axis which are in the range of $ 1.0 \le \gamma \le 8.0 $. When the initial density and total mass are constants, the major axis changes with $\gamma$. The inclination angle is kept at 45$\mdeg$. The ionization radiation penetration depth ranges from 0.225 to 0.320\% as listed in Table \ref{G1200(5)-angle-45}.

\subsubsection{Morphology of final structures}
\begin{figure*}
%\begin{minipage}{0.85\textwidth}
\includegraphics[width=0.85\textwidth]{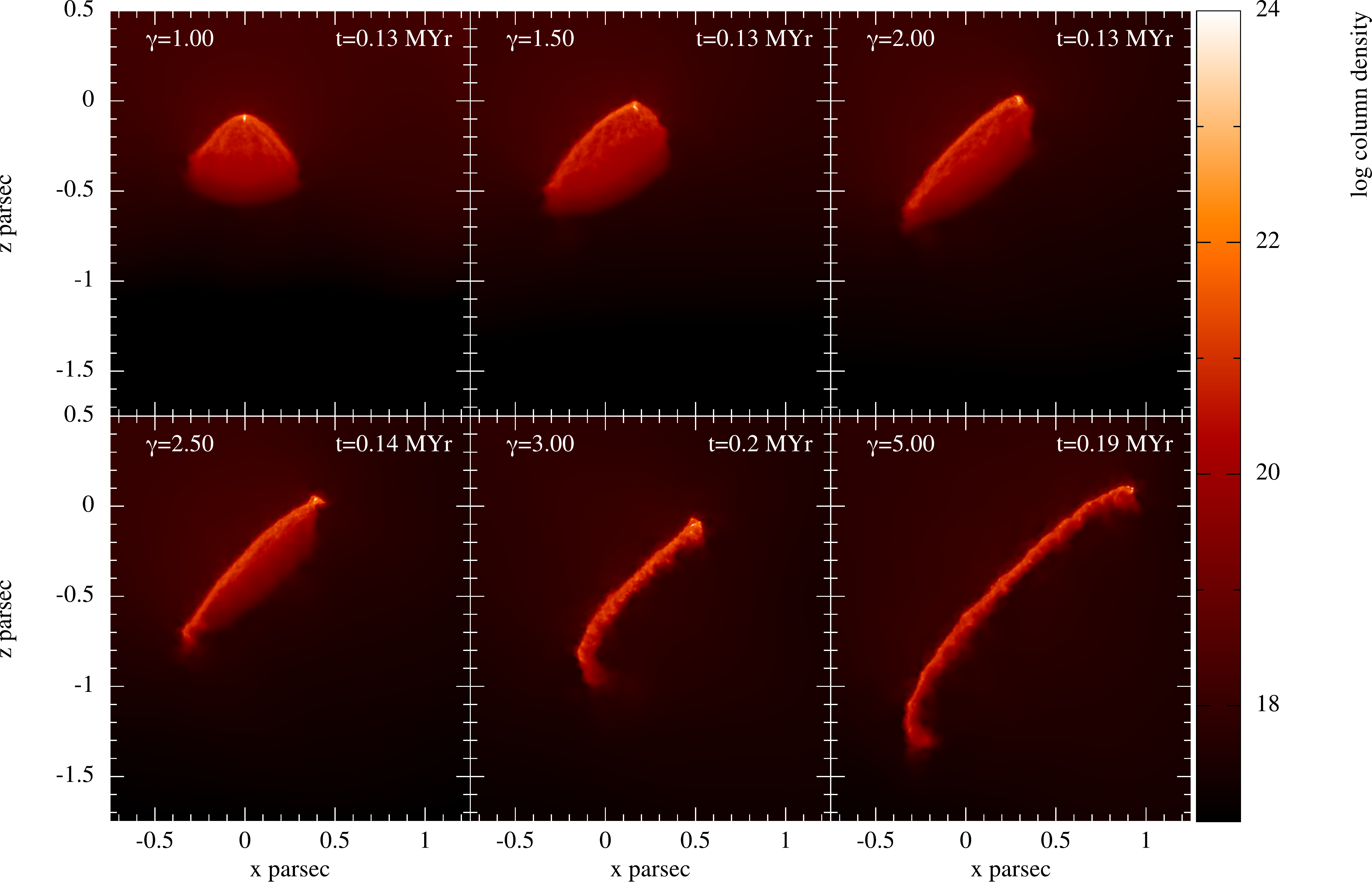}
%\end{minipage}
\caption{The final snapshots of the column density from six simulations of G1200 with $\gamma$ equal to the values shown in the top-left corner of each panel. The inclination angle is 45$\mdeg$ for all of the initial clouds.}
\label{col-den-45-all-gamma}
\end{figure*}

The dynamical evolution of these clouds is also qualitatively similar to that discussed in Section \ref{G1200_5}. We focus on the analysis of the effect of the initial shape of a cloud on the final morphology, the location(s) of the condensed core(s) and the total core masses. Figure \ref{col-den-45-all-gamma} presents six simulation result of cloud G1200 with different initial major to minor axis ratio $\gamma = $1.0, 1.5, 2.0, 2.5, 3.0, 5.0, selected from simulation results for 19 different initial clouds, described in Table \ref{G1200(5)-angle-45}. With increasing $\gamma$, the final morphology changes from a standard (symmetrical) type A ($\gamma = 1.0$), to an asymmetric type A BRC ($\gamma =1.5$), then to an asymmetric type B BRC ($\gamma = 2.0$), next to evolve to an asymmetrical type C BRC ($\gamma = 2.5$). With further increase ($\gamma \ge 3.0$) only a filamentary structure can form.

In these six simulations, high density cores form at the head of the structure except in the case of $\gamma = 2.5$, where the highly condensed core is embedded in the small `nose' structure just outside the head. It is interesting to notice that the morphological structure of SFO46 (CG1) from observations \citep{harju-1990, haikala-2010, Makela-2013} bears a similar feature to that of $\gamma = 2.5$. We leave the discussion for the formation mechanism of this morphology to the Section \ref{diff-initial-den-flux}, where we present and analyse more results of various kinds of morphological structures. It is also emphasised that we don't intend to compare the detailed physical structures between simulation and observation on SFO46 in this paper, but only to point out the link between the interesting features from these simulations to such observations. Detailed comparison of physical properties requires further simulations, and will be addressed in a subsequent paper.

\subsubsection{Displacement of the high density cores}
\begin{figure}
\center
\includegraphics[width=0.45\textwidth]{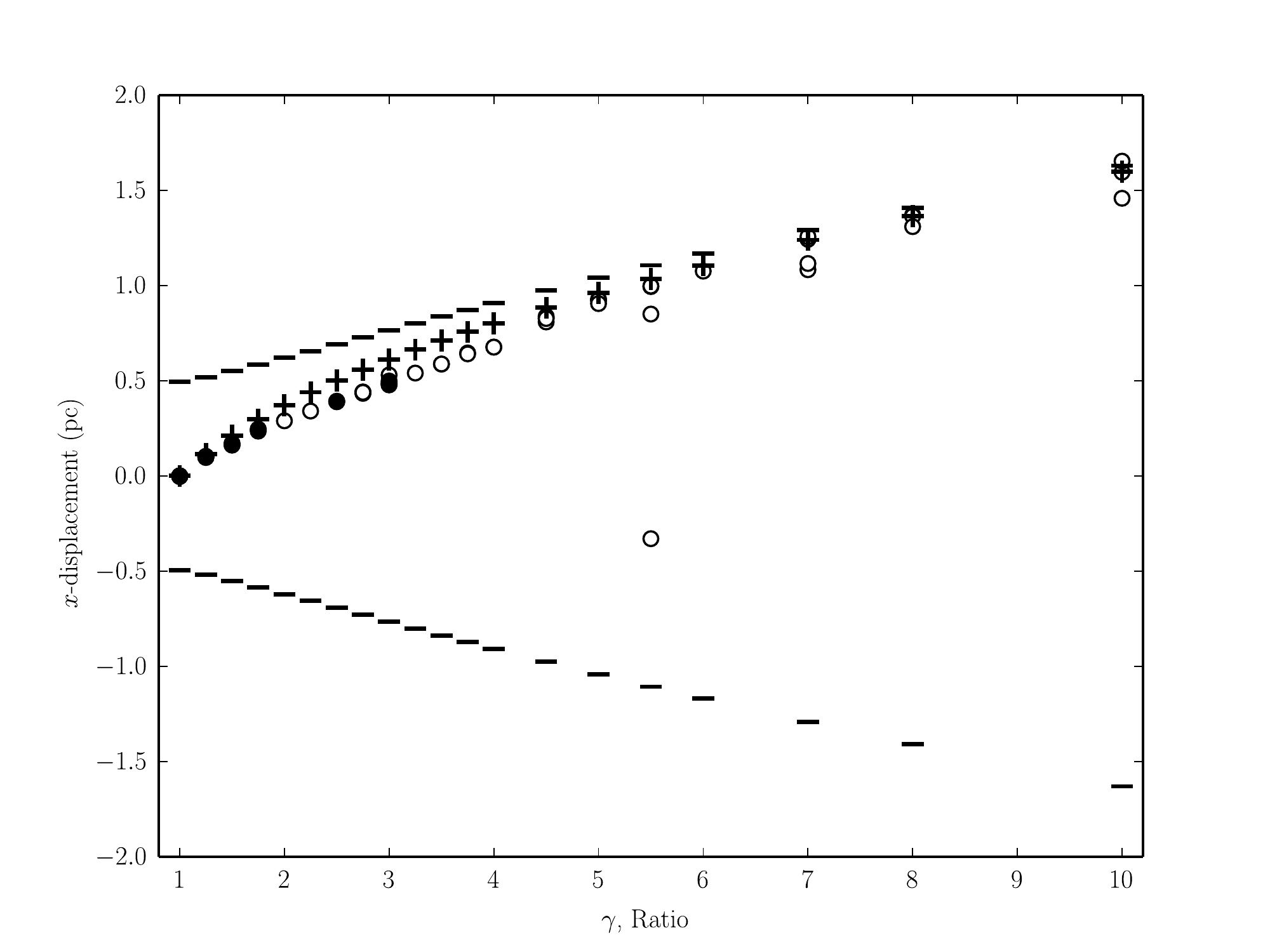}
\caption{The $x$ displacement of condensed core in G1200 series with an inclination of $\varphi = 45\%$ for varied $\gamma$. The short `-' indicates $x_{\mathrm{max}}$, `+' $x_{\mathrm{p}}$, `$\circ$' for the cores with $10^6 \le n_{\mathrm{peak}} \le 10^{12}$ \htwoden, and `$\bullet$' for the extremely high density cores with $n_{\mathrm{peak}} > 10^{12}$ \htwoden.}
\label{disp_norm_den_1200_ang_45}
\end{figure}

Figure \ref{disp_norm_den_1200_ang_45} displays the $x$-displacement of condensed core(s) with changing geometry, based on the simulation results of 19 clouds with different $\gamma$.

With increasing $\gamma$, $x_{\mathrm{p}}$ moves further in the $+x$ direction. This pattern is displayed in the same form for the condensed cores. A very clear trend of the core formation occurring at or around the apex for all ratios is exhibited.

\subsubsection{The total core mass}
\begin{figure}
\center
\includegraphics[width=0.45\textwidth]{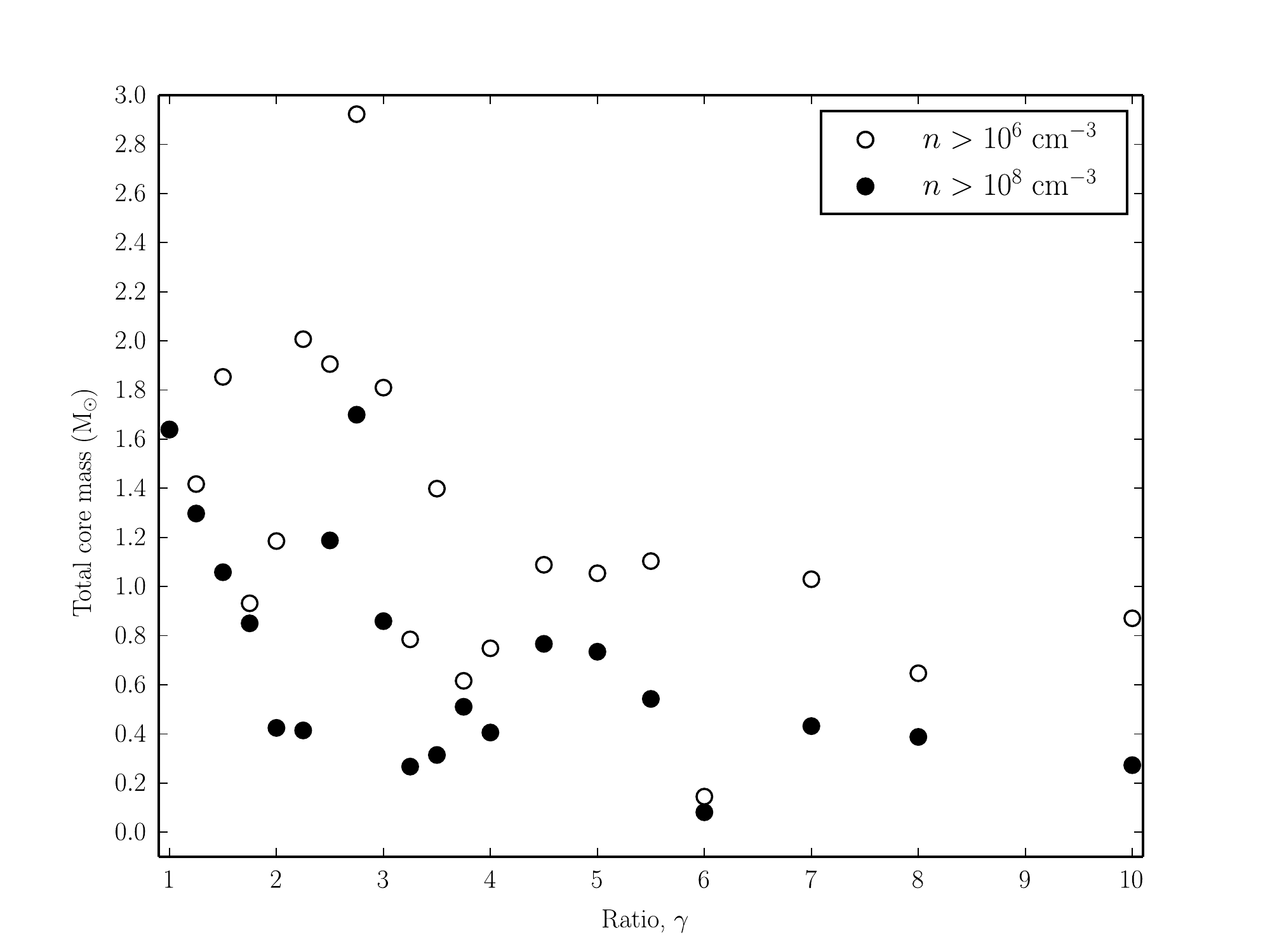}
\caption{Total mass of condensed cores in cloud G1200 series at an inclination of 45\% for varied axial ratio $\gamma$.}
\label{mass_r2_1200_all_gamma_criteria}
\end{figure}

The trend of total mass of both the high density core(s) (empty circles) and extremely high density cores (solid black circles) is a decrease with $\gamma$. This is illustrated in Figure \ref{mass_r2_1200_all_gamma_criteria}. The total mass of high density cores has an approximate pattern of maximum mass of 2.0 \msun\ at $\gamma \approx 2.25$ (with an outlier as high as almost 3.0 \msun\ at $\gamma = 2.75$) and a minimum of 0.6 \msun\ at $\gamma = 8$ (with an unusually low value of only around 0.2 \msun\ for $\gamma = 6.0$). Clouds with lower ellipticity appear to have a higher probability of collecting greater mass in the final cores. This can be expected as the mass per unit length along the major axis decreases with $\gamma$. As a result, the amount of mass able to collapse toward the apex region decreases, which is similar to the case when $\varphi = 0\mdeg$ \citep{Kinnear2014-1}. When $\varphi \ne 0\mdeg$, due to the wide range of variation in the final morphologies of the clouds, the amplitude of the fluctuation is similarly increased.

%Compare the behavior of total mass of the high density cores over $\varphi$ with the monotonic variation of high density core mass over $\varphi$, it is reasonable to think that changing the initial geometry (through $\gamma$) of a prolate cloud induces more intriguing effect on the evolution of a prolate cloud than rotaing the cloud, because in the former, not only the effective EUV radiation cross section area (see Equation \ref{eff_area}) has been changed, but also the initial gravitational state of a prolate cloud.

From the same figure, we can see the mass of extremely condensed cores (solid black circles) in a cloud also decreases with $\gamma$ in a similar way as the high density core. The mass of these cores corresponds to the precursor mass for possible proto-stars forming later in the evolution of the cloud, having a maximum of $\approx 1.6$ and a minimum of $\approx 0.2$ \msun.

\subsection{Morphological evolution of clouds of different initial density and under different ionizing flux}
\label{diff-initial-den-flux}

\begin{table}
\centering
\begin{tabular}{lcccc}
\cline{1-5}
Index & $n$  &  $a$  &  $\deuv(F_1)$ & $\deuv(F_2)$\\
&  \htwoden &  pc & \% & \% \\
G100  & 100  &  1.797  & 11.779 & 23.558 \\
G200  & 200  &  1.427  & 3.708  & 7.416  \\
G400  & 400  &  1.132  & 1.17   & 2.34   \\
G600  & 600  &  0.989  & 0.594  & 1.188  \\
G700  & 700  &  0.940  & 0.460  & 0.920  \\
G800  & 800  &  0.899  & 0.368  & 0.736  \\
G1000 & 1000 &  0.834  & 0.254  & 0.508  \\
G1200 & 1200 &  0.784  & 0.187  & 0.374  \\
\cline{1-5}
\end{tabular}
\caption{Parameters of molecular clouds of mass 30 \msun, an inclination angle $\varphi = 60\mdeg$, $\gamma =2$, and $a_{\mathrm{crit}} = 0.079$ (pc). From left to right, columns are the cloud index, initial number density, the major axis and \deuv\ defined by Equation \ref{pene-depth}, with different ionization flux $F_1$ and $F_2$ being $1.0 \times 10^{9}$ and $ 2.0 \times 10^{9}$ cm$^{-2}$ s$^{-1}$ respectively.}
\label{clouds-diff-n}
\end{table}

After the investigation of the effects of both initial shape and inclination angle of a prolate cloud on its RDI triggered dynamical evolution, we now explore the consequences of changing the initial density of a cloud and the ionizing radiation flux in the morphological development of a cloud of mass 30 \msun. This is essentially to examine the consistency of the \deuv\ parameter in characterising a wider variety of cloud parameters.

A sequence of 8 different initial densities are chosen as 100, 200, 400, 600, 700, 800, 1000, and 1200 \htwoden. Two different ionization fluxes of $F_1 = 1.0 \times 10^{9}$ and $F_2 = 2.0 \times 10^{9}$ cm$^{-2}$ s$^{-1}$ are applied for a total of 16 simulations.
An inclination angle of $60\mdeg$ and $\gamma = 2.0$ is used for all cases. This is as a result of the simulation of G1200(5) exhibiting the formation of the `nose' structure in the vicinity of these geometries.
As such, we can also expect to get some clue of how the `nose' structure is developed and changes with the initial properties of a cloud in this set of simulations. The corresponding parameters of the clouds are listed in Table \ref{clouds-diff-n}. It is seen that the change in EUV radiation penetration depth over the initial density ($100 \le n \le 1200$ \htwoden) is almost two orders of magnitude, a much more dramatic variance than from changing $\gamma$ or $\varphi$.

\begin{figure*}
%\begin{minipage}{1.0\textwidth}
\includegraphics[width=0.95\textwidth]{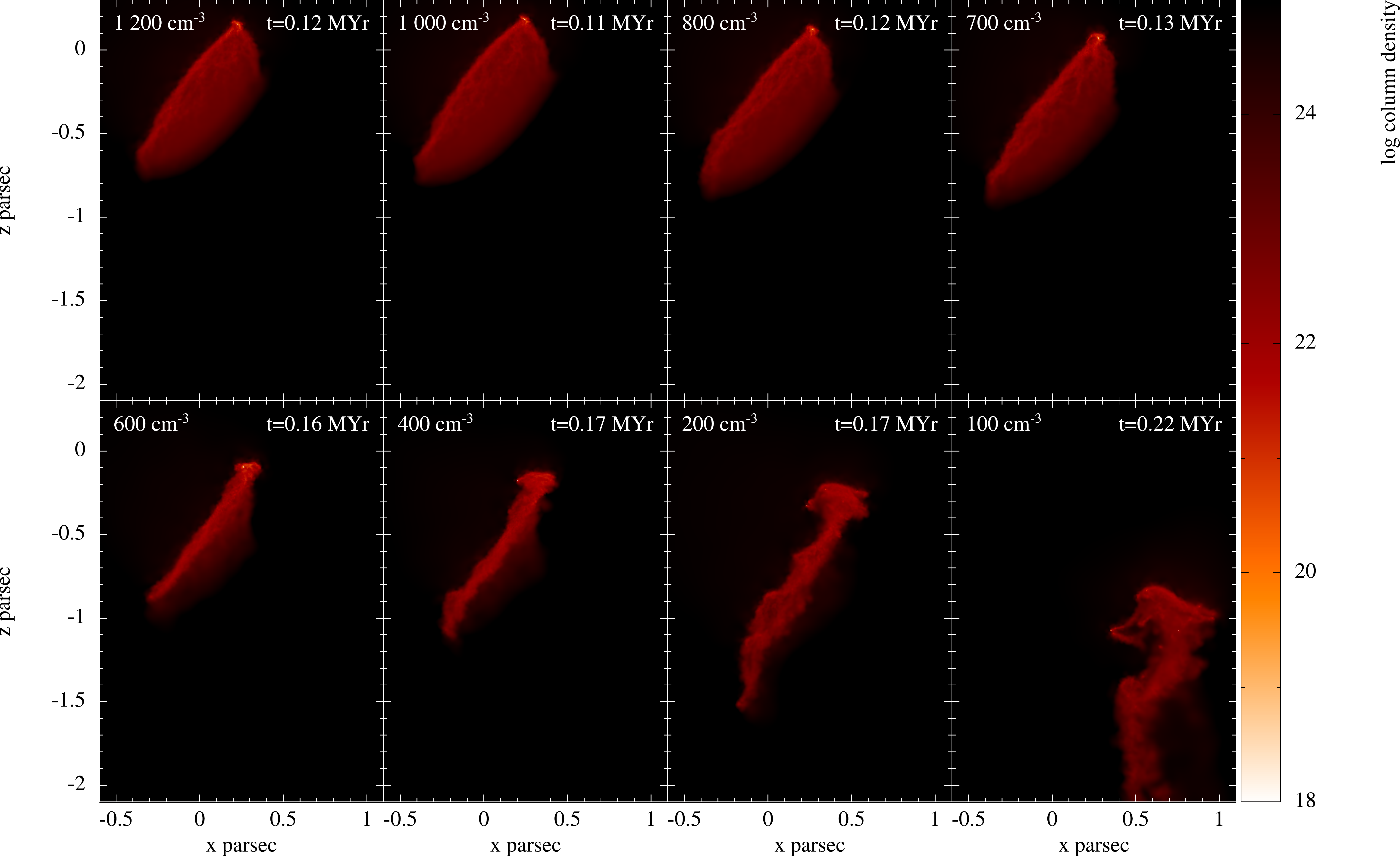}
%\end{minipage}
\caption{The final snapshot of the column density of clouds of initial mass of 30 \msun\ $\gamma = 2.0$ and $\varphi = 60 \mdeg$\ but different initial densities as shown in the top-left corner of each panel. The ionization flux = 10$^{9}$ cm$^{-2}$ s$^{-1}$}
\label{density_differences_flux_1e9}
\end{figure*}

\begin{figure*}
%\begin{minipage}{1.0\textwidth}
\includegraphics[width=0.95\textwidth]{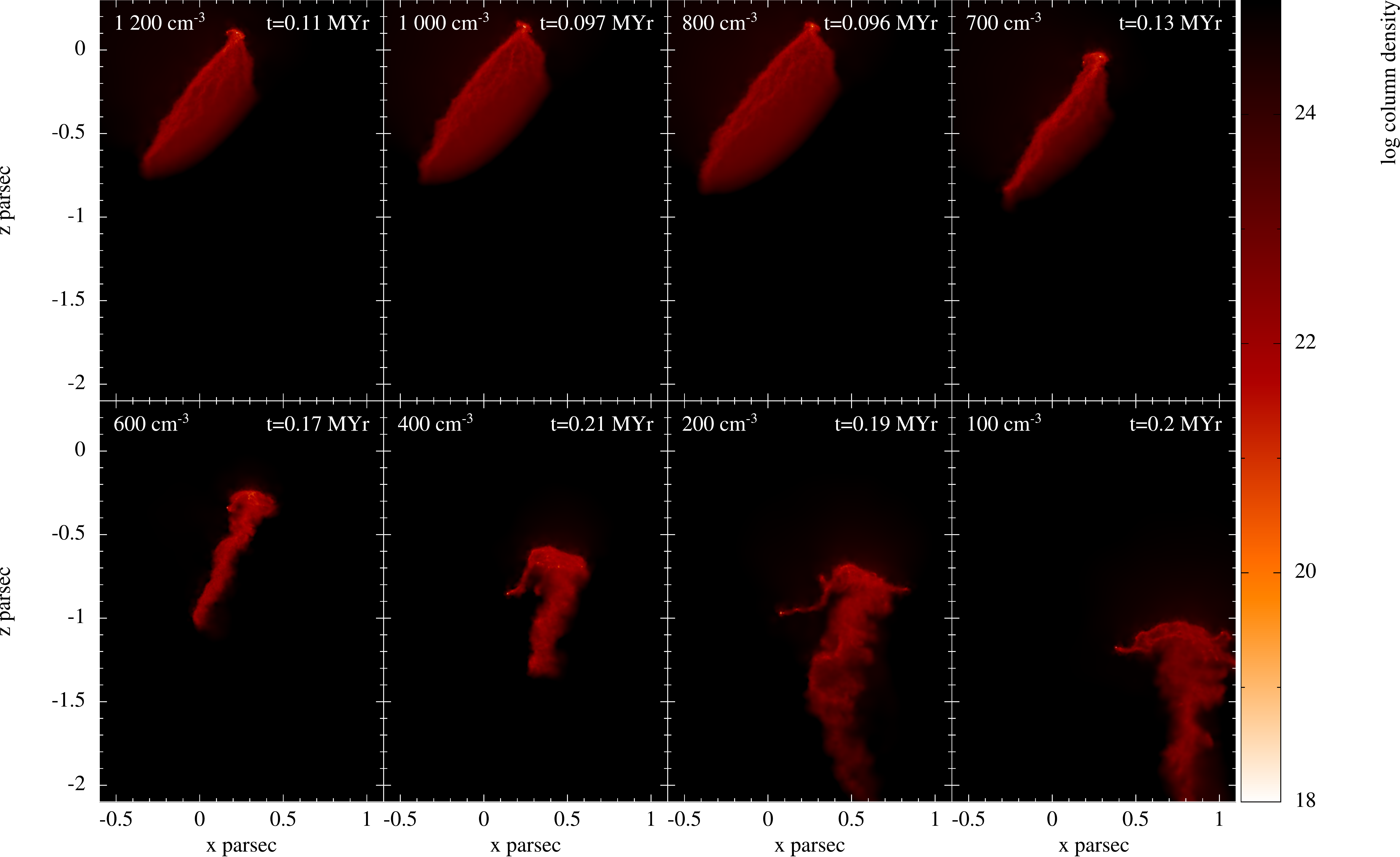}
%\end{minipage}
\caption{The final snapshot of the column density of the same 8 clouds as in Figure \ref{density_differences_flux_1e9} (initial mass of 30 \msun\ $\gamma = 2.0$ and $\varphi = 60 \mdeg$) but with increased incident flux of $2 \times 10^{9}$ cm$^{-2}$ s$^{-1}$.}
\label{density_differences_flux_2e9}
\end{figure*}

\subsubsection{An overview of the final morphological structures}
Figure \ref{density_differences_flux_1e9} displays the final snapshots of the column density of the 8 simulations with ionization flux of $F_1$. The morphology of the final structure of a cloud changes from an asymmetrical type C BRC to a filamentary structure and then to an irregular structure, as the initial density is decreased from 1200 to 100 \htwoden, which leads to an increase of \deuv\ from 0.187 to 11.779\%.

As shown in the top row panels, an asymmetrical type C BRC is developed in G1200, G1000, G800, and G700, with \deuv\ of 0.187, 0.254, 0.368 and 0.460\% respectively. Although they all bear an asymmetric type C BRC morphology, the minor axis becomes narrower with increasing \deuv, which results in more cloud material being photo-evaporated from their star facing surfaces. When the initial density $n = 600$ \htwoden\ ($\deuv = 0.594$\%), the prolate cloud evolves into a filamentary structure, as shown in the first panel in the bottom row of Figure \ref{density_differences_flux_1e9}. Clouds of initial densities of 400, 200 and 100 \htwoden\ (\deuv\ = 1.17, 3.708 and 11.779 \% respectively) are all seen to form irregular structures.

\begin{figure}
\center
\includegraphics[width=0.45\textwidth]{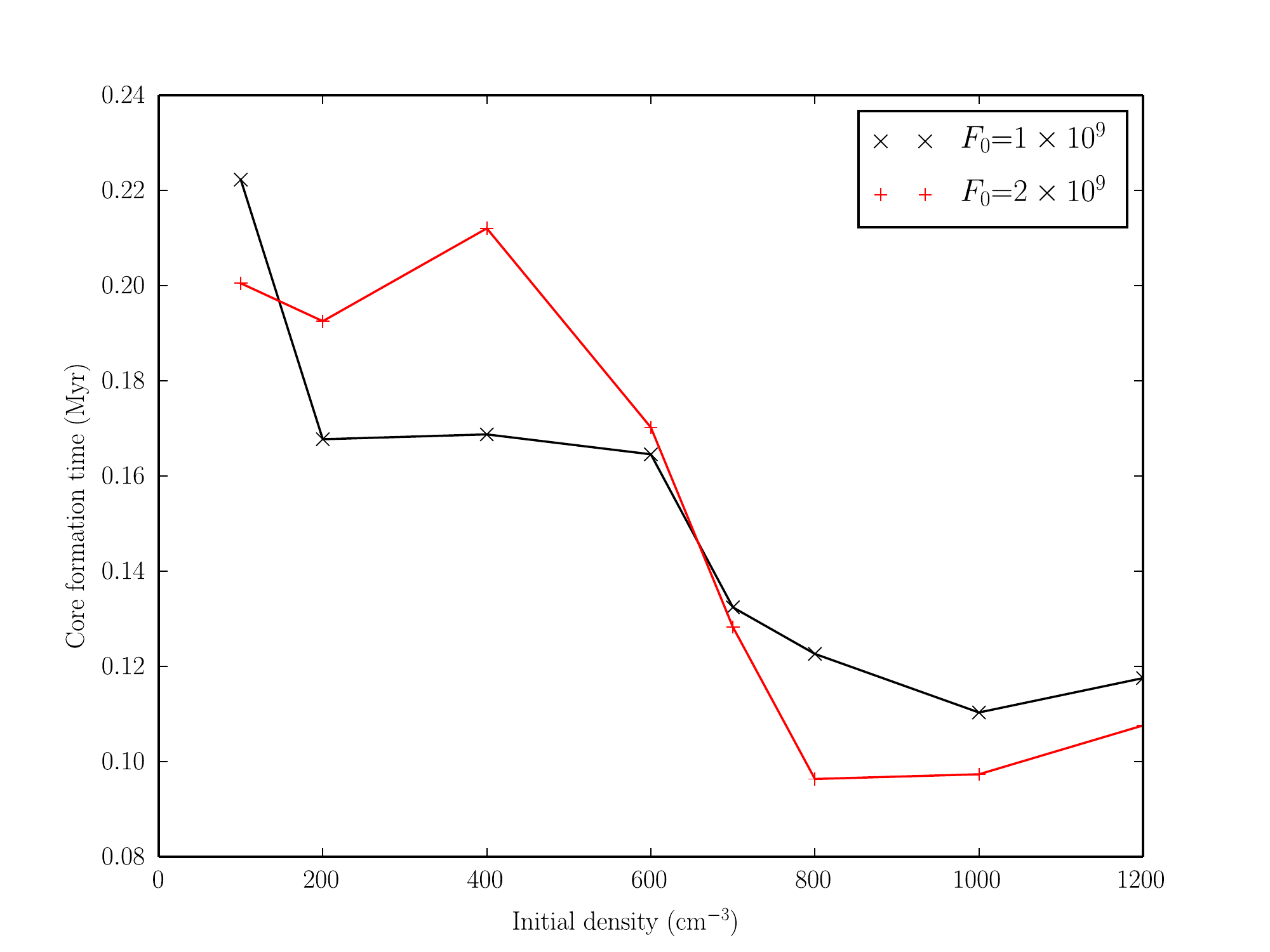}
\caption{The core formation time of the $F_1$ and $F_2$ group clouds vs their initial densities.}
\label{init_den_vs_core_time}
\end{figure}

\begin{figure}
\center
\includegraphics[width=0.45\textwidth]{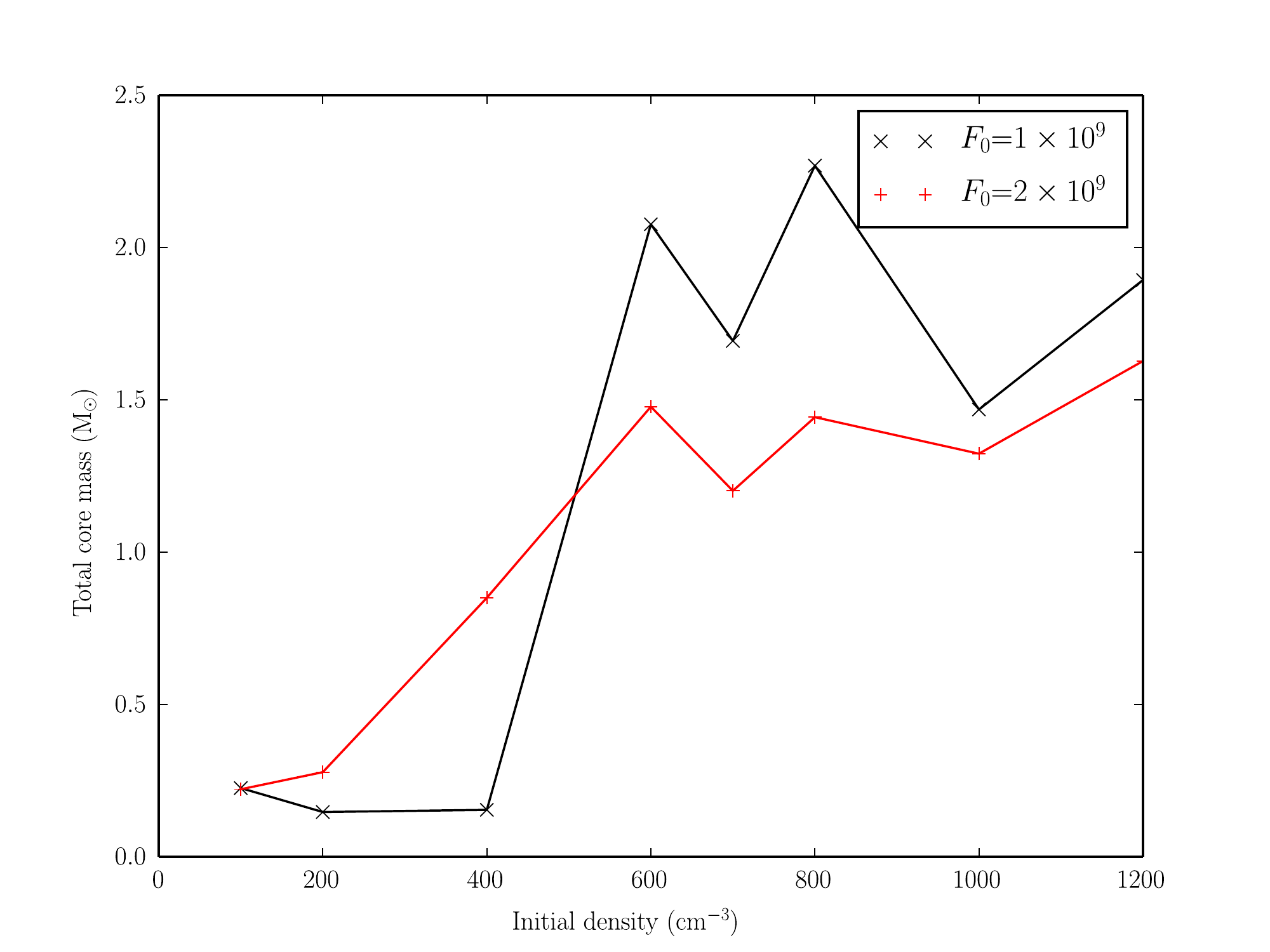}
\caption{The total mass of the cores formed in $F_1$ and $F_2$ group clouds vs their initial densities.}
\label{core_mass_50_002_1e6}
\end{figure}

The final morphological structures from the simulations using doubled ionization flux $F_2$ are presented in Figure \ref{density_differences_flux_2e9}. Similar morphological structure sequences to those seen in Figure \ref{density_differences_flux_1e9} are observed. An asymmetrical type C BRC is formed from G1200, G1000 and G800, a filamentary structure from G700, and irregular structures in G600, G400, G200 and G100. The only difference is that the morphological transition point shifts to a cloud of higher initial density compared to the $F_1$ simulations.

The first transition from type C BRC to filamentary morphology occurs around G600($F_1$) and G700($F_2$), and the second transition from filamentary to irregular morphology is around G400($F_1$) and G600($F_2$). These appear to correspond to similar \deuv\ in each case, for both fluxes. The former transition occurs at $\deuv \approx 0.8$ and the latter at $\deuv \approx 1.1$. This indicates that \deuv\ remains a characteristic parameter for the dynamical evolution of a cloud for the parameter space so far investigated.

Figures \ref{init_den_vs_core_time} and \ref{core_mass_50_002_1e6} present the core formation time and the total mass of the RDI triggered condensed cores vs the initial density of the clouds of the above two groups of simulated clouds, with black lines for $F_1$ group and red for $F_2$. It is shown that the core formation time in both group clouds decreases with the initial density of a cloud. This is because of the stronger initial gravitational bonding in the high density cloud. For the same reason, the total mass of the condensed cores follow an increasing trend with the initial density of the cloud. However the effect of the strength of the EUV radiation flux on the efficiency of triggered core formation vs initial density is two fold. When the initial density is below 600 \htwoden\ the clouds illuminated by the higher flux ($F_2$) take a longer time to form condensed cores, and collect slightly more mass in their condensed cores than in the clouds illuminated by the lower flux ($F_1$). This trend is reversed when the initial density is higher then 600 \htwoden. i.e., clouds illuminated by higher flux take less time to form condensed core and collect less mass into their condensed cores than the clouds illuminated by lower flux.

Overall, the information delivered in the above two figures suggests that the RDI triggered star formation efficiency is much higher (shorter core formation time and higher total core mass) in molecular clouds of shorter \deuv\ (corresponding to higher initial density) in both $F_1$ and $F_2$ group clouds.%, or say in regular BRCs structures, than in irregular structures which have longer \deuv\ (correspoint to lower initial density). We can understand this feature easily because EUV radiation irradiates more gas material from the acting clouds which has a deeper EUV radiation penetration depth.

From the above two sets of simulations, we can see that when $\deuv \ge 1.0\%$, the deep penetration of the EUV radiation into the cloud causes irregular structure formation. A closer look at the above two morphological variation sequences tells that with increasing \deuv, the tiny `nose' structure around the apex of the asymmetrical type C BRC gradually grows and becomes an irregular `horse-head' structure. In order to understand how the `nose' and `horse-head' structures form, we further investigate the evolutionary process of one of the above clouds, which presents a clear horse-head-like morphology.

\subsubsection{Growing from `nose' to `horse-head' structure}

\begin{figure*}
%\begin{minipage}{0.85\textwidth}
%\includegraphics[width=0.50\textwidth]{column-evo-den400-60-2.pdf}
\includegraphics[width=0.70\textwidth]{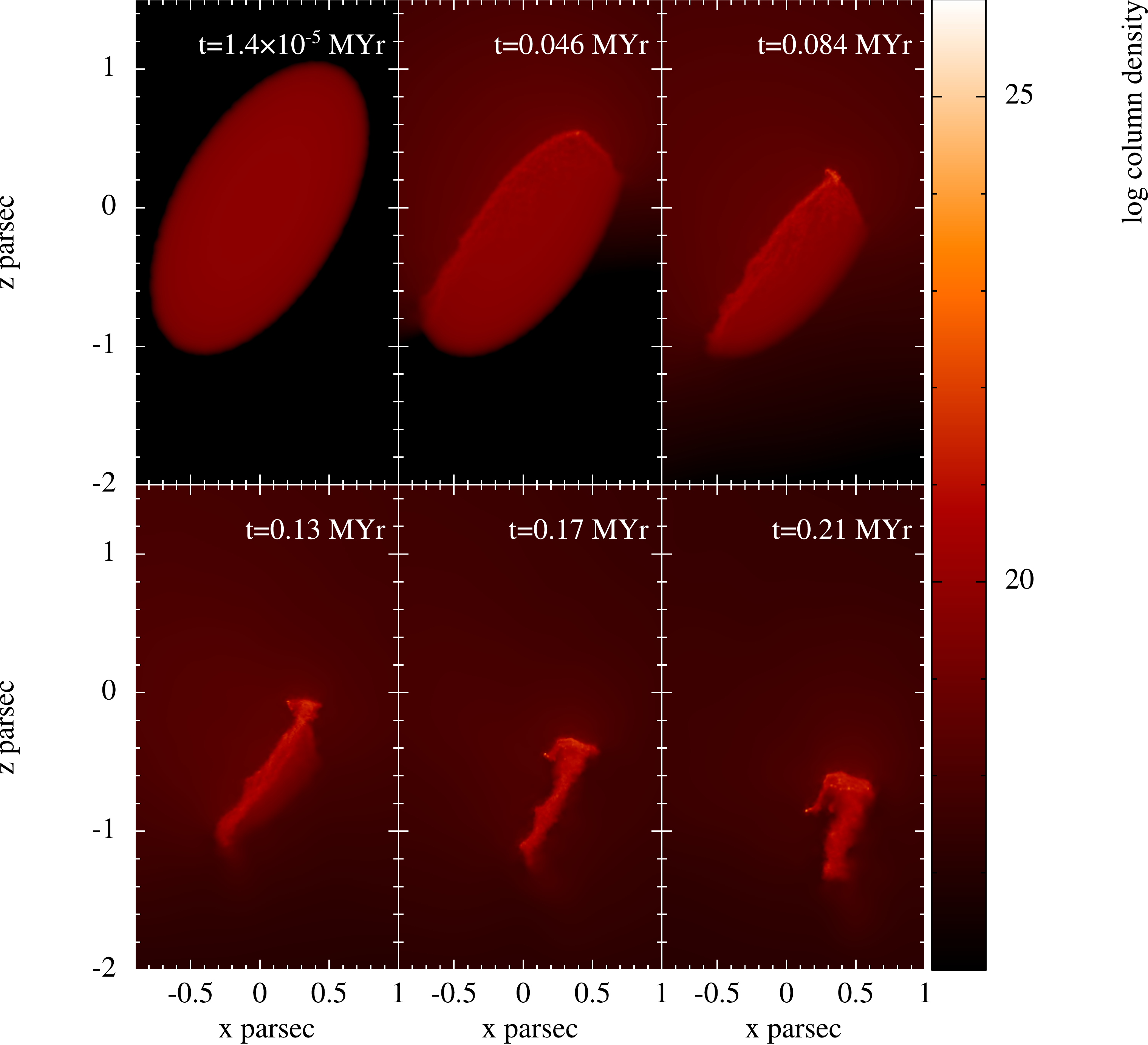}
%\end{minipage}
\caption{The column density evolution of a cloud G400($F_2$), with an inclination angle of 60$\mdeg$.}
\label{column-den400}
\end{figure*}

\begin{figure*}
%\begin{minipage}{0.85\textwidth}
%\includegraphics[width=0.65\textwidth]{velocity_scaled_filtered.pdf}
\includegraphics[width=0.70\textwidth]{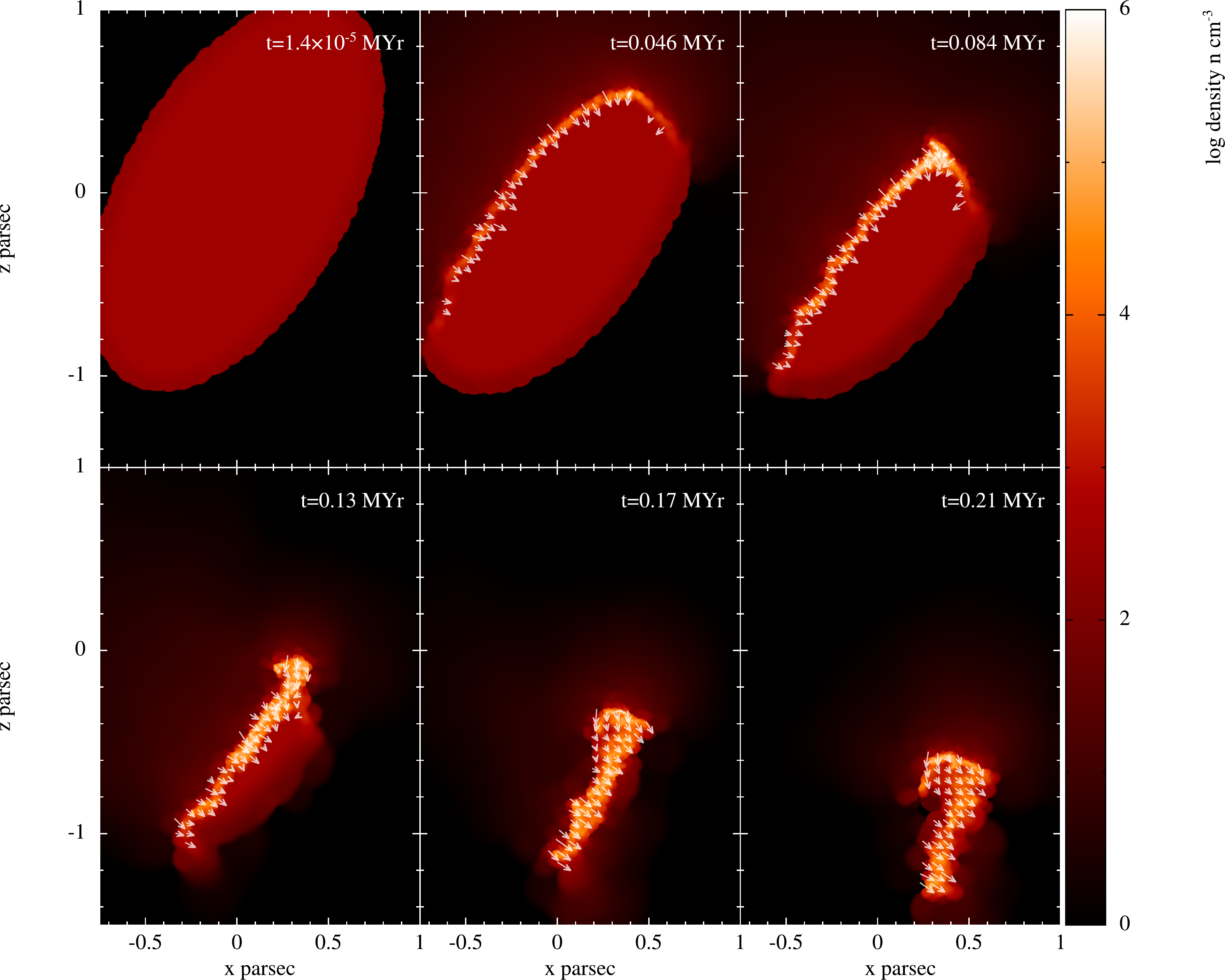}
%\end{minipage}
\caption{The evolution of the RDI induced shock velocity of cloud G400($F_2$) overlapped in the corresponding cross-section density snapshot. The magnitude of the velocity is between 2 and 12 km s$^{-1}$.}
\label{velocity-evolution}
\end{figure*}

We chose the cloud G400($F_2$) to analyse its evolutionary process due to its formation of a particularly distinctive `horse-head' structure.
Figure \ref{column-den400} describes a six-snapshot column density evolutionary sequence of cloud G400($F_2$), which shows a growing process from a small `nose' to a `horse-head' morphological structure. The corresponding RDI induced shock velocity profiles for each of the snapshots are plotted in Figure \ref{velocity-evolution}, which reveals the kinematic mechanism for the morphology formation in G400($F_2$).

From Figure \ref{column-den400}, we can see that a shocked layer forms at $t = 0.046$ Myr, which surrounds most of the surface of the front semi-ellipsoid and a small part of the top surface of the back semi-ellipsoid. The maximum magnitude of the shock velocity $v_{\mathrm{s}} = 12$ km $s^{-1}$, calculated as the peak velocity of non-ionized gas within the condensed shock front. Most of the shocked layer surrounding the front semi-ellipsoid propagates into the cloud smoothly toward (+$x$, -$z$) direction. Very close to the apex, the shock velocity of the front semi-ellipsoid is toward the direction of ($x, -z$), and that of the back semi-ellipsoid is ($-x,-z$), the gas material from the two sides of the apex converge beneath the apex to make a small  `nose' structure  at $t = 0.084$ Myr,  as shown in the top-right panels Figures \ref{column-den400} and \ref{velocity-evolution}. 

While a small `nose' structure forms at the point of ($x$, $z$) $\approx$ (0.2, 0.2) pc and with $v_{\mathrm{x}} \approx 0$, below the nose structure, the RDI induced shock keeps pushing the surface layer of the front semi-ellipsoid moves toward the ($+x$, $-z$) direction at the shock velocity. This makes the below-nose region of the BRC gradually produce a displacement along $x$ direction relative to the nose. The nose becomes a short horizontal fragment at $t = 0.13$ Myr, in which the direction of the RDI induced shock velocity is mainly along $z$ direction. The short horizontal fragment gradually forms a `hook' at $t = 0.17$ Myr as shown in the lower middle panels of Figures \ref{column-den400} and \ref{velocity-evolution}. Further propagation of the shocked layer below the hook toward ($+x$, $-z$) direction makes the `hook' structure grow to look like a `horse-head' as shown in the last panels in the above two figures. The final `horse-head' structure bears some resemblance to the well known horse-head nebula.

It is seen from the above analysis that the appearance of a `horse-head' morphology in this instance is a consequence of growing a small `nose' structure. The `nose' structure forms if the gas around the apex of an ellipsoidal cloud converge beneath that apex due to a strong RDI induced shock effect, as described above. This requires that the cloud should have a reasonably high inclination angle, i.e., $\varphi > 45\mdeg$ in our simulations. However the formation of a `nose' structure does not necessarily lead to a `horse-head' structure. For example, a small `nose' formation is seen from the final morphologies of G700-1200 in the $F_1$ series and G800-1200 in the $F_2$ series. Whether the `nose' can grow into a `horse-head' also depends on the initial ionization penetration depth, i.e., $\deuv > 1.0$\%. The latter is to guarantee a higher enough shock velocity $\left(v_{\mathrm{s}} \sim \sqrt{\frac{F_{\mathrm{Lyman}}}{n}}\right)$\citep{Bertoldi1989-1}, which can push the majority of the gas below the nose toward the ($+x$, $-z$) direction.

\subsection{Development of other irregular morphologies}
\label{diff-morphology}
Further simulations were conducted, to explore if there are other structures which form different irregular morphologies to a `horse-head' as a consequence of the variation of initial conditions of the cloud and ionization flux. A series of 16 simulations with G400($F_2$) of different $\gamma = 2.0, 2.5, 3.0, 3.5$ and $\varphi = 60, 70, 80, 85\mdeg$ are conducted. For all 16 clouds, $\deuv > 1.0$\%. In this section, we concentrate on the description of the morphological variations, as the underlying mechanisms appear to be similar, and they are of interest due to their variety rather than exact evolutions.

\begin{figure*}
\includegraphics[width=0.95\textwidth]{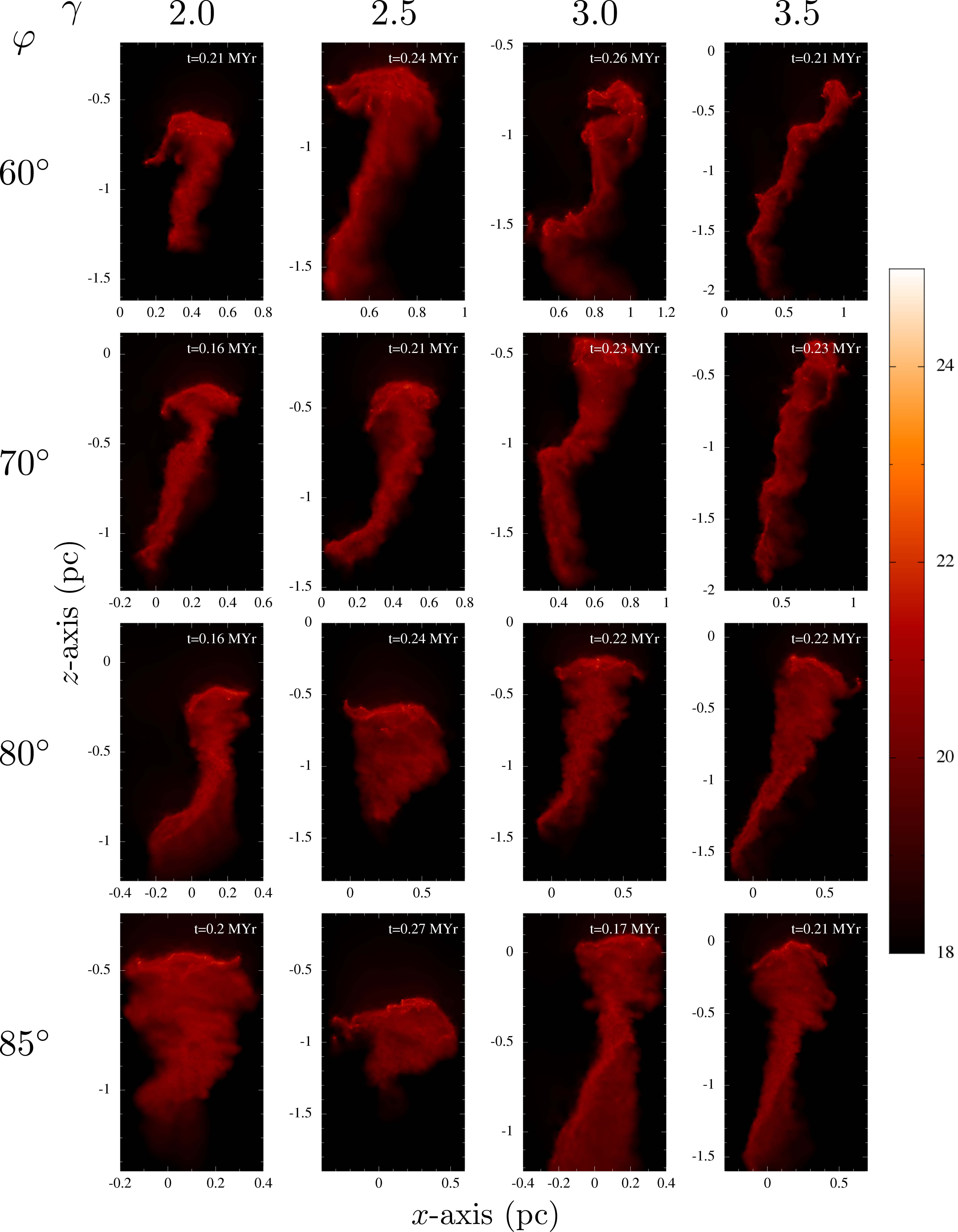}
\caption{The final morphologies of the clouds G400($F_2$) of different initial geometry and inclination angle. The axis scales vary between panels, because of the different initial scale of the major- and minor axes for the clouds depicted.}
\label{diff-morphology-N400}
\end{figure*}

The first column panels in Figure \ref{diff-morphology-N400} present the final morphologies of G400($F_2$) of $\gamma = 2$ and 4 different initial inclination angles, which show that when the initial angle $\varphi$ increases from 60$\mdeg$, the `horse-head' in the final structure becomes longer in $z$ direction. The connection point between the `horse-head' and the below-horse-head part moves downward with $\varphi$. At 85$^{\mdeg}$, the whole morphology is an `elephant trunk' like structure. This is because the `nose' structure formed in the first tens of Kyr of the evolution increasingly stretches along the $z$ axis with increasing $\varphi$. The first row panels in Figure \ref{diff-morphology-N400} describe the morphology change of cloud G400($F_2$) of inclination angle $\varphi =60\mdeg$ and varied $\gamma$. When the initial prolate cloud becomes more elliptical, in addition to a `horse-head' structure, the RDI induced shock also triggers linear gravitational instability along the direction of the major axis in some clouds.

The 16 panels in Figure \ref{diff-morphology-N400} present 16 different final morphologies which appear in the clouds of the initial ($\gamma, \varphi$) as stated above. We can conclude that the larger the inclination angle, the longer the `horse-head' sub-structure; and that the more elliptical the cloud initially, the thinner the final morphology is. Although, as in previous instances, their structural details differ across the range of $\gamma$.

Through viewing more simulation results obtained using different combination of ($F_{\mathrm{Lyman}}, n, \gamma, \varphi$), we find that the variety of the final morphology is extensive for $\deuv > 1.0$\%, although it is impossible to show all of them in this paper.

\section{Conclusions}

The results from the simulations conducted show that uniform density prolate clouds at an inclination to a plane parallel EUV radiation field can evolve to a variety of morphological structures often found at various \HII\ boundaries. The parameter of EUV radiation penetration depth, \deuv, of a given cloud can be a good indicator as to whether it would evolve to a regular BRC or irregular structures.

The final morphology of the clouds of initial denisty ($1200$ \htwoden) and ellipticity ($\gamma = 2$), and with $\deuv < 1.0$\% changes from a filament to an asymmetrical BRC, then to a symmetrical BRC, as the inclination angle $\varphi$ increases from $0$ to $90\mdeg$. The core formation time and the total mass of the RDI triggered condensed cores both decrease with $\varphi$. Therefore EUV radiation triggered star formation is quicker in high $\varphi$ prolate clouds, but the total mass of the triggered stars is higher for low $\varphi$ cloud.

For clouds of fixed initial density ($1200$ \htwoden) and inclination angle ($\varphi = 45\mdeg$), as $\gamma$ increases from 1 to 8, the final morphologies  changes from a symmetrical to asymmetrical BRC, then to filamentary structure. RDI triggered star formation efficiency decreases with the increase of $\gamma$, due to a greater quantity of gas material being evaporated with a high cross sectional area to the incident radiation.

It is found that the final morphology of a prolate cloud is very sensitive to the initial density. For fixed initial shape ($\gamma = 2.0$) and fixed inclination angle ($\varphi = 60 \mdeg$), the morphology of a cloud changes from an asymmetric type C BRCs to a filament then to irregular structures, such as a `horse-head', as the initial density  decreases from 1200 to 100 \htwoden. EUV radiation triggered star formation efficiency is found to be higher in regular BRCs (with $\deuv < 1.0$\%) than in irregular structures (with $\deuv > 1.0$\%).

%%%%%%In general, stronger EUV fluxes induce higher `low density' core formation rates in low starting density clouds, meaning a greater number of potential cores but each with low mass and undeveloped collapse. Weaker EUV fluxes are more efficient at triggering high density cores in high density environments. As expected, however, high density clouds will, regardless of flux, have a higher chance of permitting triggered high density cores.

%A stronger EUV radiation field is slightly more efficient to RDI triggered very low mass core formation in clouds of lower initial density, than a weaker EUV radiation field does, the latter is more efficent in triggering high density core formation in cloud of higher initial density. However the RDI triggered core formation time is much shorter and the total mass of the condense core is much higher in clouds of high initial density than in that of lower initial density, when illuminated by both ionization radiation flux.

Based on the above systematic investigation on the evolution of molecular clouds (or clumps) at an \HII\ boundary,
we can now suggest a unified formation mechanism for the different morphological structures seen in the published images of a variety of \HII\ regions: Giant molecular clouds (GMCs) are clumpy and the numerous clumps are of different shapes, from spherical to highly elliptical, and with their major axes not exhibiting a strong preference for any particular alignment within their environment. When a star or a cluster of stars forms inside a GMC, the interaction between the ionizing radiation with the surrounding molecular clumps of different shapes and orientations  creates structures of various morphologies. The spherical and prolate clumps aligned with the radiation flux form standard type BRCs, the inclined prolate clumps with a shallow ionizing radiation penetration depth form asymmetrical BRCs or an inclined linear structures; the inclined prolate clumps with a deep radiation penetration depth form a variety of irregular structures, the well known  `horse-head' structure is only one of them.

In our next paper, we are going to investigate the evolution of physical properties of a few observed asymmetrical BRCs and the RDI triggered star formation inside them.
\section{Acknowledgement}
Timothy Kinnear thanks the University of Kent for providing his PhD studentship.
\bibliography{paper_two_TMK}

\begin{thebibliography}{}

\bibitem[\protect\citeauthoryear{{Bastien}}{{Bastien}}{1983}]{Bastien1983-1}
{Bastien} P.,  1983, A\&A, 119, 109

\bibitem[\protect\citeauthoryear{{Bate} \& {Burkert}}{{Bate} \&
  {Burkert}}{1997}]{BateBurkert1997-1}
{Bate} M.~R.,  {Burkert} A.,  1997, MNRAS, 288, 1060

\bibitem[\protect\citeauthoryear{{Bertoldi}}{{Bertoldi}}{1989}]{Bertoldi1989-1}
{Bertoldi} F.,  1989, ApJ, 346, 735

\bibitem[\protect\citeauthoryear{{Bisbas}, {W{\"u}nsch}, {Whitworth}, {Hubber}
  \& {Walch}}{{Bisbas} et~al.}{2011}]{BisbasEtAl2011-1}
{Bisbas} T.~G.,  {W{\"u}nsch} R.,  {Whitworth} A.~P.,  {Hubber} D.~A.,
  {Walch} S.,  2011, ApJ, 736, 142

\bibitem[\protect\citeauthoryear{{Chauhan}, {Pandey}, {Ogura}, {Jose}, {Ojha},
  {Samal} \& {Mito}}{{Chauhan} et~al.}{2011}]{ChauhanEtAl2011-1}
{Chauhan} N.,  {Pandey} A.~K.,  {Ogura} K.,  {Jose} J.,  {Ojha} D.~K.,  {Samal}
  M.~R.,    {Mito} H.,  2011, MNRAS, 415, 1202

\bibitem[\protect\citeauthoryear{{Doty}, {Everett}, {Shirley}, {Evans} \&
  {Palotti}}{{Doty} et~al.}{2005}]{Doty-2005}
{Doty} S.~D.,  {Everett} S.~E.,  {Shirley} Y.~L.,  {Evans} N.~J.,    {Palotti}
  M.~L.,  2005, MNRAS, 359, 228

\bibitem[\protect\citeauthoryear{{Dyson} \& {Williams}}{{Dyson} \&
  {Williams}}{1997}]{DysonWilliams1997-1}
{Dyson} J.~E.,  {Williams} D.~A.,  1997, {The physics of the interstellar
  medium}.
Bristol: Institute of Physics Publishing

\bibitem[\protect\citeauthoryear{{Fukuda}, {Miao}, {Sugitani}, {Kawahara},
  {Watanabe}, {Nakano} \& {Pickles}}{{Fukuda} et~al.}{2013}]{FukudaEtAl2013-1}
{Fukuda} N.,  {Miao} J.,  {Sugitani} K.,  {Kawahara} K.,  {Watanabe} M.,
  {Nakano} M.,    {Pickles} A.~J.,  2013, ApJ, 773, 132

\bibitem[\protect\citeauthoryear{{Gammie}, {Lin}, {Stone} \&
  {Ostriker}}{{Gammie} et~al.}{2003}]{Gammie-2003}
{Gammie} C.~F.,  {Lin} Y.-T.,  {Stone} J.~M.,    {Ostriker} E.~C.,  2003, ApJ,
  592, 203

\bibitem[\protect\citeauthoryear{{Gholipour} \& {Nejad-Asghar}}{{Gholipour} \&
  {Nejad-Asghar}}{2013}]{Gholipour-2013}
{Gholipour} M.,  {Nejad-Asghar} M.,  2013, MNRAS, 429, 3166

\bibitem[\protect\citeauthoryear{{Gong} \& {Ostriker}}{{Gong} \&
  {Ostriker}}{2011}]{Hao-2011}
{Gong} H.,  {Ostriker} E.~C.,  2011, ApJ, 729, 120

\bibitem[\protect\citeauthoryear{{Gritschneder}, {Naab}, {Walch}, {Burkert} \&
  {Heitsch}}{{Gritschneder} et~al.}{2009}]{Gritschneder2009-1}
{Gritschneder} M.,  {Naab} T.,  {Walch} S.,  {Burkert} A.,    {Heitsch} F.,
  2009, ApJL, 694, L26

\bibitem[\protect\citeauthoryear{{Habing}}{{Habing}}{1968}]{Habing1968-1}
{Habing} H.~J.,  1968, Bull.~Astron.~Inst.~Netherlands, 19, 421

\bibitem[\protect\citeauthoryear{{Haikala}, {M{\"a}kel{\"a}} \&
  {V{\"a}is{\"a}nen}}{{Haikala} et~al.}{2010}]{haikala-2010}
{Haikala} L.~K.,  {M{\"a}kel{\"a}} M.~M.,    {V{\"a}is{\"a}nen} P.,  2010,
  A\&A, 522, A106

\bibitem[\protect\citeauthoryear{{Harju}, {Sahu}, {Henkel}, {Wilson}, {Sahu} \&
  {Pottasch}}{{Harju} et~al.}{1990}]{harju-1990}
{Harju} J.,  {Sahu} M.,  {Henkel} C.,  {Wilson} T.~L.,  {Sahu} K.~C.,
  {Pottasch} S.~R.,  1990, A\&A, 233, 197

\bibitem[\protect\citeauthoryear{{Haworth} \& {Harries}}{{Haworth} \&
  {Harries}}{2012}]{HaworthHarries2012-1}
{Haworth} T.~J.,  {Harries} T.~J.,  2012, MNRAS, 420, 562

\bibitem[\protect\citeauthoryear{{Karr}, {Noriega-Crespo} \& {Martin}}{{Karr}
  et~al.}{2005}]{Karr-2005}
{Karr} J.~L.,  {Noriega-Crespo} A.,    {Martin} P.~G.,  2005, Aj, 129, 954

\bibitem[\protect\citeauthoryear{{Kessel-Deynet} \& {Burkert}}{{Kessel-Deynet}
  \& {Burkert}}{2000}]{KesselBurkert2000-1}
{Kessel-Deynet} O.,  {Burkert} A.,  2000, MNRAS, 315, 713

\bibitem[\protect\citeauthoryear{{Kessel-Deynet} \& {Burkert}}{{Kessel-Deynet}
  \& {Burkert}}{2003}]{KesselBurkert2003-1}
{Kessel-Deynet} O.,  {Burkert} A.,  2003, MNRAS, 338, 545

\bibitem[\protect\citeauthoryear{{Kinnear}, {Miao}, {White} \&
  {Goodwin}}{{Kinnear} et~al.}{2014}]{Kinnear2014-1}
{Kinnear} T.~M.,  {Miao} J.,  {White} G.~J.,    {Goodwin} S.,  2014, MNRAS,
  444, 1221

\bibitem[\protect\citeauthoryear{{Kusune}, {Sugitani}, {Miao}, {Tamura},
  {Sato}, {Kwon}, {Watanabe}, {Nishiyama}, {Nagayama} \& {Sato}}{{Kusune}
  et~al.}{2014}]{Kusune2014-1}
{Kusune} T.,  {Sugitani} K.,  {Miao} J.,  {Tamura} M.,  {Sato} Y.,  {Kwon} J.,
  {Watanabe} M.,  {Nishiyama} S.,  {Nagayama} T.,    {Sato} S.,  2014, ArXiv
  e-prints

\bibitem[\protect\citeauthoryear{{Lefloch} \& {Lazareff}}{{Lefloch} \&
  {Lazareff}}{1994}]{LeflochLazareff1994-1}
{Lefloch} B.,  {Lazareff} B.,  1994, A\&A, 289, 559

\bibitem[\protect\citeauthoryear{{M{\"a}kel{\"a}} \&
  {Haikala}}{{M{\"a}kel{\"a}} \& {Haikala}}{2013}]{Makela-2013}
{M{\"a}kel{\"a}} M.~M.,  {Haikala} L.~K.,  2013, A\&A, 550, A83

\bibitem[\protect\citeauthoryear{{Miao}, {Sugitani}, {White} \&
  {Nelson}}{{Miao} et~al.}{2010}]{MiaoEtAl2010-1}
{Miao} J.,  {Sugitani} K.,  {White} G.~J.,    {Nelson} R.~P.,  2010, ApJ, 717,
  658

\bibitem[\protect\citeauthoryear{{Miao}, {White}, {Nelson}, {Thompson} \&
  {Morgan}}{{Miao} et~al.}{2006}]{MiaoEtAl2006-1}
{Miao} J.,  {White} G.~J.,  {Nelson} R.,  {Thompson} M.,    {Morgan} L.,  2006,
  MNRAS, 369, 143

\bibitem[\protect\citeauthoryear{{Miao}, {White}, {Thompson} \&
  {Nelson}}{{Miao} et~al.}{2009}]{MiaoEtAl2009-1}
{Miao} J.,  {White} G.~J.,  {Thompson} M.~A.,    {Nelson} R.~P.,  2009, ApJ,
  692, 382

\bibitem[\protect\citeauthoryear{{Morgan}, {Thompson}, {Urquhart}, {White} \&
  {Miao}}{{Morgan} et~al.}{2004}]{MorganEtAl2004-1}
{Morgan} L.~K.,  {Thompson} M.~A.,  {Urquhart} J.~S.,  {White} G.~J.,    {Miao}
  J.,  2004, A\&A, 426, 535

\bibitem[\protect\citeauthoryear{{Nelson} \& {Langer}}{{Nelson} \&
  {Langer}}{1999}]{NelsonLanger1999-1}
{Nelson} R.~P.,  {Langer} W.~D.,  1999, ApJ, 524, 923

\bibitem[\protect\citeauthoryear{{Nelson} \& {Papaloizou}}{{Nelson} \&
  {Papaloizou}}{1993}]{Nelson-1993}
{Nelson} R.~P.,  {Papaloizou} J.~C.~B.,  1993, MNRAS, 265, 905

\bibitem[\protect\citeauthoryear{{Oort} \& {Spitzer} Jr.}{{Oort} \&
  {Spitzer}}{1955}]{OortSpitzer1955-1}
{Oort} J.~H.,  {Spitzer} Jr. L.,  1955, ApJ, 121, 6

\bibitem[\protect\citeauthoryear{{Osterbrock}}{{Osterbrock}}{1957}]{Osterbrock-1957}
{Osterbrock} D.~E.,  1957, ApJ, 125, 622

\bibitem[\protect\citeauthoryear{{Panwar}, {Chen}, {Pandey}, {Samal}, {Ogura},
  {Ojha}, {Jose} \& {Bhatt}}{{Panwar} et~al.}{2014}]{Panwar-2014}
{Panwar} N.,  {Chen} W.~P.,  {Pandey} A.~K.,  {Samal} M.~R.,  {Ogura} K.,
  {Ojha} D.~K.,  {Jose} J.,    {Bhatt} B.~C.,  2014, ArXiv e-prints

\bibitem[\protect\citeauthoryear{{Price}}{{Price}}{2007}]{Splash}
{Price} D.~J.,  2007, PASA, 24, 159

\bibitem[\protect\citeauthoryear{{Rathborne}, {Johnson}, {Jackson}, {Shah} \&
  {Simon}}{{Rathborne} et~al.}{2009}]{RathborneEtAl2009-1}
{Rathborne} J.~M.,  {Johnson} A.~M.,  {Jackson} J.~M.,  {Shah} R.~Y.,
  {Simon} R.,  2009, ApJS, 182, 131

\bibitem[\protect\citeauthoryear{{Sicilia-Aguilar}, {Roccatagliata}, {Getman},
  {Henning}, {Mer{\'{\i}}n}, {Eiroa}, {Rivi{\`e}re-Marichalar} \&
  {Currie}}{{Sicilia-Aguilar} et~al.}{2014}]{Sicilia-Aguilar-2014}
{Sicilia-Aguilar} A.,  {Roccatagliata} V.,  {Getman} K.,  {Henning} T.,
  {Mer{\'{\i}}n} B.,  {Eiroa} C.,  {Rivi{\`e}re-Marichalar} P.,    {Currie} T.,
   2014, A\&A, 562, A131

\bibitem[\protect\citeauthoryear{{Springel}}{{Springel}}{2005}]{Springel2005-1}
{Springel} V.,  2005, MNRAS, 364, 1105

\bibitem[\protect\citeauthoryear{{Sugitani}, {Fukui} \& {Ogura}}{{Sugitani}
  et~al.}{1991}]{SugitaniEtAl1991-1}
{Sugitani} K.,  {Fukui} Y.,    {Ogura} K.,  1991, ApJs, 77, 59

\bibitem[\protect\citeauthoryear{{Sugitani} \& {Ogura}}{{Sugitani} \&
  {Ogura}}{1994}]{SugitaniOgura1994-1}
{Sugitani} K.,  {Ogura} K.,  1994, ApJS, 92, 163

\bibitem[\protect\citeauthoryear{{Thompson}, {Urquhart} \& {White}}{{Thompson}
  et~al.}{2004}]{Thompson-2004}
{Thompson} M.~A.,  {Urquhart} J.~S.,    {White} G.~J.,  2004, A\&A, 415, 627

\bibitem[\protect\citeauthoryear{{Tremblin}, {Audit}, {Minier} \&
  {Schneider}}{{Tremblin} et~al.}{2012}]{Tremblin-2012}
{Tremblin} P.,  {Audit} E.,  {Minier} V.,    {Schneider} N.,  2012, A\&A, 538,
  A31

\bibitem[\protect\citeauthoryear{{Urquhart}, {Thompson}, {Morgan} \&
  {White}}{{Urquhart} et~al.}{2006}]{UrquhartEtAl2006-1}
{Urquhart} J.~S.,  {Thompson} M.~A.,  {Morgan} L.~K.,    {White} G.~J.,  2006,
  A\&A, 450, 625

\end{thebibliography}
\bibliographystyle{mn2e}

\end{document}